\documentclass[11pt]{article}

\usepackage{verbatim}
\usepackage{amsmath}
\usepackage{amsfonts}
\usepackage{amssymb}
\usepackage{bbm}
\usepackage{latexsym}
\usepackage{lscape} 
\usepackage{epsfig}
\usepackage{color}
\usepackage{graphicx}
\usepackage[table]{xcolor}
\usepackage{enumerate}
\usepackage{hhline}
\usepackage{subfig}
\usepackage{multirow}
\usepackage{mathrsfs}
\usepackage{longtable}

\usepackage{amscd}
\usepackage{cite}

\renewcommand{\arraystretch}{1.5} 

\DeclareMathSymbol{\mg}{\mathrel}{symbols}{"1D}

\newcommand{\ga}{\alpha}

\newcommand{\gd}{\delta}
\renewcommand{\ge}{\epsilon}

\newcommand{\gk}{\kappa}

\newcommand{\gL}{\Lambda}

\newcommand{\cN}{{\cal N}}

\newcommand{\dsp}{\displaystyle}

\newcommand{\beq}{\begin{equation}}
\newcommand{\eeq}{\end{equation}}
\newcommand{\barr}{\begin{array}}
\newcommand{\earr}{\end{array}}
\newcommand{\equ}[1]{\begin{gather} #1 \end{gather}}

\newcommand{\items}[1]{\begin{itemize} #1 \end{itemize}}

\newcommand{\sfrac}[2]{\mbox{$\frac{#1}{#2}$}}
\newcounter{oldcounter}

\newcommand{\Intr}{\mathbb{Z}}

\newcommand{\ba}[2]{\[\begin{array}{#2}\label{#1}}
\newcommand{\ea}{\end{array}\]}
\newcommand{\be}{\begin{equation}}
\newcommand{\ee}{\end{equation}}
\newcommand{\bea}{\begin{eqnarray}}
\newcommand{\eea}{\end{eqnarray}}

\newcommand{\rep}[1]{\mathbf{#1}}
\newcommand{\crep}[1]{\overline{\rep{#1}}}
\newcommand{\brep}[1]{\overline{\rep{#1}}}

\newcommand{\sm}{{\,\mbox{-}}}

\newcommand{\arrystrch}{1.2}

\usepackage{hyperref} 
\usepackage{cancel}

\begin{document}

\thispagestyle{empty}

\begin{flushright}
DESY-15-126, LMU-ASC 47/15
\\
\end{flushright}
\begin{center}
{\Large {\bf 
(MS)SM-like models on smooth Calabi-Yau manifolds from all three heterotic string theories
} 
}
\\[0pt]

\bigskip
\bigskip 
{\large
{\bf{Stefan Groot Nibbelink}$^{a,}$\footnote{E-mail: Groot.Nibbelink@physik.uni-muenchen.de}},  
{\bf{Orestis Loukas}$^{a,b,}$\footnote{E-mail: O.Loukas@physik.uni-muenchen.de}},
{\bf{Fabian Ruehle$^{c,}$}\footnote{E-mail: fabian.ruehle@desy.de}}
\bigskip}\\[0pt]
\vspace{0.23cm}
${}^a$ {\it 
Arnold Sommerfeld Center for Theoretical Physics,   \\ 
Ludwig-Maximilians-Universit\"at M\"unchen, 80333 M\"unchen, Germany
}
\\[1ex]
${}^b$ {\it School of Electrical and Computer Engineering, \\
National Technical University of Athens, Zografou Campus, GR-15780 Athens, Greece}
\\[1ex]
${}^c$ {\it 
Deutsches Elektronen-Synchrotron DESY, Notkestrasse 85, 22607 Hamburg, Germany
}
\\[1ex] 
\bigskip
\end{center}

\subsection*{\centering Abstract}

We perform model searches on smooth Calabi-Yau compactifications for both the supersymmetric E$_8\times$E$_8$ and SO(32) as well as for the non-supersymmetric SO(16)$\times$SO(16) heterotic strings simultaneously. 
We consider line bundle backgrounds on both favorable CICYs with relatively small $h_{11}$ and the Schoen manifold. 
Using Gram matrices we systematically analyze the combined consequences of the Bianchi identities and the tree-level Donaldson-Uhlenbeck-Yau equations inside the K\"ahler cone. 
In order to evaluate the model building potential of the three heterotic theories on the various geometries, we perform computer-aided scans. 
We have generated a large number of GUT-like models (up to over a few hundred thousand on the various geometries for the three heterotic theories) which become (MS)SM-like upon using a freely acting Wilson line. 
For all three heterotic theories we present tables and figures summarizing the potentially phenomenologically interesting models which were obtained during our model scans.

\newpage 
\setcounter{page}{1}
 \setcounter{footnote}{0}
\tableofcontents
\newpage

\section{Introduction}
\label{sc:Introduction}

There exist three heterotic string theories in ten dimensions that have modular invariant partition functions and are tachyon-free: the supersymmetric E$_8\times$E$_8$ and SO(32) theories and the non-supersymmetric SO(16)$\times$SO(16) theory. Even though all three of them were roughly discovered around the same time, essentially only the E$_8\times$E$_8$ theory has been examined in string model building on Calabi-Yau manifolds in a more systematic fashion. On smooth compact orbifold resolutions semi-realistic models were obtained from the E$_8\times$E$_8$ theory using line bundles~\cite{Nibbelink:2009sp,Blaszczyk:2010db,Nibbelink:2012de}. 
In the class of favorable Complete Intersection Calabi-Yaus (CICYs) with discrete symmetries, specific S(U(1)$^5$) line bundle constructions embedded in a single E$_8$ factor have been investigated in~\cite{Anderson:2011ns,Anderson:2012yf,Anderson:2013xka}.

On the other hand, the SO(32) theory has only been studied sporadically regarding phenomenological applications, although the general framework has been laid out in~\cite{Blumenhagen:2005zg}. 
The model building potential of the non-supersymmetric SO(16)$\times$SO(16) theory compactified on smooth Calabi-Yau geometries has not really been investigated so far; in~\cite{Font:2002pq,Blaszczyk:2014qoa} this option was at least mentioned. In particular, as explained in \cite{Blaszczyk:2014qoa,Blaszczyk:2015zta}, one can avoid tachyons to leading order in $\ga^\prime$ and $g_s$ when working in the large volume regime on smooth Calabi-Yaus.

Systematic model scans for MSSM (Minimal supersymmetric Standard Model)-like models have again been performed for the E$_8\times$E$_8$ theory on  string backgrounds that admit exact CFT (Conformal Field Theory) descriptions. The most prominent ones have been toroidal orbifold model searches: In Refs.~\cite{Buchmuller:2005jr,Buchmuller:2006ik,Lebedev:2006kn,Lebedev:2008un} MSSM--like models have been constructed on the ${\mathbb Z}_\text{6--II}$ orbifold. Related investigations have been carried out on the $\Intr_2\times \Intr_4$ orbifold~\cite{Z2xZ4}, the $\Intr_\text{12--I}$ orbifold~\cite{Kim:2006hv,Kim:2007mt}, a $\Intr_2\times\Intr_2$ orbifold~\cite{Blaszczyk:2009in} and $\Intr_8$ orbifolds~\cite{Nibbelink:2013lua}; an overview can be found in Ref.~\cite{Nilles:2014owa}. In addition, there have been model scans using the free fermionic formulation 
\cite{Cleaver:1998sa,Faraggi:2007tj,Faraggi:2014hqa,Faraggi:2014vma}. 
Moreover, also Gepner constructions have been used for model searches~\cite{Dijkstra:2004ym,Dijkstra:2004cc}. 

Model searches for non-supersymmetric string constructions with an exact CFT description have been sporadically considered in the past compared to those in the supersymmetric context. A pioneering investigation of non-supersymmetric models was carried out in Ref.~\cite{Rohm:1983aq}, which proposes for the first time ways to break supersymmetry spontaneously in string theory (see also \cite{Kounnas:1989dk}). After that it was shown that it is possible to obtain constructions that interpolate between the supersymmetric heterotic E$_8\times$E$_8$ and then non-supersymmetric SO(16)$\times$SO(16) theories~\cite{Itoyama:1986ei}. Conditions for vacuum stability in non-supersymmetric closed string models were determined in~Ref.~\cite{Kutasov:1990sv}: In order to avoid tachyons the total number of (massless and massive) bosonic and fermionic degrees of freedom must be equal asymptotically. Refs.~\cite{Dienes:1994np,Dienes:2006ut} pointed out that the spectrum fulfills certain supertrace relations which are very similar to supersymmetric theories, and therefore this was called misaligned supersymmetry. The corresponding oscillatory behavior of the (classically stable) non-supersymmetric closed strings spectra are related to the non-trivial zeros of the Riemann zeta function~\cite{Angelantonj:2010ic}. Non-supersymmetric orbifolds were investigated in Refs.~\cite{Toon:1990ij,Font:2002pq}. Further aspects of non-supersymmetric models in heterotic and other string contexts can be found e.g.\ in~\cite{Harvey:1998rc,Aldazabal:1999jr,Antoniadis:1999xk,Angelantonj:1999gm,Angelantonj:1999ms,Dudas:2000ff,Dudas:2002dg,Angelantonj:2007ts}. Non-supersymmetric models were constructed from rational CFTs in~Refs.~\cite{GatoRivera:2007yi,GatoRivera:2008zn}.

Moreover, recently such investigations have been carried out by various groups in the heterotic orbifold constructions~\cite{Blaszczyk:2014qoa} and free fermionic models~\cite{Abel:2015oxa,Ashfaque:2015vta}. In~\cite{Angelantonj:2014dia} it was found that the difference of thresholds for non-abelian gauge still exhibits a universal behavior in some heterotic constructions with spontaneously broken supersymmetry.

Stimulated by this revived interest in non-supersymmetric string constructions one of the main purposes of this work is to investigate the phenomenological prospects of the non-supersymmetric SO(16)$\times$SO(16) string on smooth Calabi-Yau spaces. The techniques that are used in this paper have been laid out in detail in the paper~\cite{Blaszczyk:2015zta}. To make the presentation  self-contained we review some of that material in this work. However, here we always have the automated implementations of these methods in mind.

At the same time, the current paper aims to present a comprehensive study of all three heterotic theories simultaneously to examine whether such a bias in favor of  the supersymmetric E$_8\times$E$_8$ heterotic string for smooth compactifications in the past is justified. For that we choose a subset of the often considered favorable CICYs that admit freely acting $\Intr_N$ symmetries and the Schoen manifold. We set up our analysis in such a way that we are able to treat the three heterotic string theories side-by-side.

To this end we consider line bundle backgrounds on these smooth geometries characterized by so-called line bundle vectors. Except for the fact that for the three different theories we need to take these vectors out of different lattices, their consistency conditions 
are identical and the resulting spectra can be computed employing very similar methods. Using these universal algorithms we can investigate the model building potential of the three heterotic theories. Since in the present work we are mainly interested in statistical findings, we try to avoid as much as possible putting any phenomenological bias in the construction of smooth models. This enables us to examine how the phenomenologically interesting models are distributed among the more general landscape of a given geometry.

To obtain systematic model building results it is almost unavoidable to make use of computer-automated model scans. In this paper we lay out in detail how we have set up such model searches using procedures that allow for quite general line bundle ansatzes, i.e.\ parameterizations of many embeddings in the whole Cartan sub-algebra of the ten-dimensional gauge group. Nevertheless our algorithms are very fast, especially due to the use of index theorems to calculate the chiral spectrum. The latter point is very important if one wants to investigate the physical consequences of many different line bundles in a statistical manner.

Since the construction and analysis of such smooth models is technically rather involved, we apply various consistency checks on our resulting spectra at the chiral level. In particular, for each constructed model we check that all charged massless states have integral multiplicities and that all (pure and mixed) gauge and gravitational anomalies cancel in four dimensions. In addition, for the big subset of the constructed models that are chiral exact (i.e.\ give exactly three chiral generations of quarks and leptons) we have also calculated the spectra using bundle cohomology. These tests provide strong cross checks on both the spectrum and the topological data entering its computation.

The model searches that we present are mainly focused on obtaining chiral GUT-like models which become MSSM-like (for the supersymmetric theories) or SM(Standard Model)-like (for the non-supersymmetric theory) upon acting with a freely acting Wilson line. 
Our side-by-side approach for these three heterotic theories means that, aside from many MSSM-like models from the E$_8\times$E$_8$ theory, we obtain a big novel set of MSSM-like constructions from the SO(32) theory. In addition, we construct a large collection of SM-like models from the non-supersymmetric SO(16)$\times$SO(16) theory. 

\subsection*{Paper overview}

The remainder of this paper is organized as follows: In Section~\ref{sc:ExamplesCYs} we briefly review the basic topological properties of the smooth Calabi-Yau manifolds we need for our model building analysis focusing on favorable CICYs with relatively small $h_{11}$ and the Schoen manifold.
In the next section we summarize an efficient parameterization of line bundles on smooth spaces. In particular we introduce a convenient rewriting of the Bianchi identities and the Donaldson-Uhlenbeck-Yau (DUY) equations. Section~\ref{sc:RandomModels} explains our model building constructions and analysis in detail. In Section~\ref{sc:Models} we implemented these ideas to perform model searches for GUT- and (MS)SM-like spectra of the three heterotic string theories. We compare the outcome of these model scans side-by-side and present statistics on four-dimensional SM chirality among models with an SU(5) gauge group.
In addition, we inspect the non-chiral spectrum at the GUT level for those (MS)SM-like models which have exactly three chiral generations of quarks and leptons.

\section{Examples of smooth Calabi-Yau manifolds}
\label{sc:ExamplesCYs} 

In this section we describe the basic topological properties of two types of Calabi-Yau manifolds which we have used in our model searches. This data contains the Hodge numbers $(h_{11}, h_{21})$ counting the number of K\"ahler and complex structure deformations, and values of the integrated second and third Chern classes
\equ{
c_{2\,i} = \int_{D_i} \widehat{c}_2~, 
\qquad 
c_3  = \int_X \widehat{c}_3~,  
}  
for $i=1,\ldots, h_{11}$.
Finally, the intersection number of three divisors $D_i, D_j$ and $D_k$ is defined by 
\equ{
\gk_{ijk} = \int_X \widehat D_i\, \widehat D_j\, \widehat D_k~, 
}
where $\widehat D_i$ denotes the (1,1)-form Poincar\'e-dual to $D_i$. In particular, the volumes of any curve $C$, divisor $D$ and the manifold $X$ itself can be obtained from
\equ{ \label{VolumesWithDilaton} 
\text{Vol}(C) = \int_C J~, 
\qquad 
 \text{Vol}(D) = \sfrac 12 \int_D  J^2~, 
\qquad 
\text{Vol}(X) = \sfrac 16 \int_X  J^3 ~, 
}
using the intersection numbers and the expansion
\equ{
J = a_i\, \widehat D_i
} 
of the K\"ahler form. Inside the Mori and the K\"ahler cones the volumes of all curves $C$ and all divisors $D$ are positive. 

Generically, the quantity
\equ{ \label{CurveIntersections} 
q_j(C) = \int_{C} \widehat{D}_j  = \int_{D_i} \widehat{C}  = \int_X \widehat{C}\, \widehat{D}_j  
}
gives the number of intersections of a curve $C$ with divisor $D_i$\,. 
If we have an integral basis of curves $\{C_i\}$ and divisors $\{D_i\}$, then 
\equ{ \label{MinIntBasis} 
q_j(C_i) = \int_{C_i} \widehat{D}_j  = \int_{D_i} \widehat{C}_j  = \int_X \widehat{C}_i \widehat{D}_j  = \gd_{ij}~. 
}
This basis is minimal when any effective curve $C$ can be written as a formal sum of the curves $C_i$ with non-negative coefficients and a similar expansion holds for any divisor $D$. In a minimal integral basis the K\"ahler/Mori cone is simply given by $a_i > 0$ for all $i=1,\ldots, h_{11}$\,. 

\subsection{Complete intersection Calabi-Yaus} 
\label{sc:CICYs}

\begin{table}
\begin{center}
\renewcommand{\arraystretch}{1.3}
\begin{tabular}{|c||c|c|}
\hline
{\bf CICY \#}						&	{\bf Total volume} 									&	{\bf Chern classes} 
\\
$(h_{11}, h_{21})$			&	$6\, \text{Vol}(X) = \kappa_{ijk}\, a_i a_j a_k$	&		$(c_{2i}\,; c_3)$
\\ \hline\hline
{\bf 7862}		&	\multirow{2}{*}{$12 a_1 a_2 a_3 + 12 a_1 a_2 a_4 + 12 a_1 a_3 a_4 + 12 a_2 a_3 a_4$}	       &	\multirow{2}{*}{$(24,24,24,24;-128)$}
\\
$(4,68)$						&   	&	
\\ \hline
{\bf 7491,\,7522}	&	$12 a_1 a_2 a_3 + 24 a_1 a_2 a_4 + 24 a_1 a_3 a_4 + 24 a_2 a_3 a_4 $ &	\multirow{2}{*}{$(24,24,24,64;-80)$}
\\
$(4,44)$          			   &	$+ 24 a_1 a_4^2 + 24 a_2 a_4^2 + 24 a_3 a_4^2 + 16 a_4^3$	&
\\ \hline
{\bf 7447,\,7487}	&	$12 a_1 a_2 a_3 + 12 a_1 a_2 a_4 + 12 a_1 a_3 a_4 + 12 a_2 a_3 a_4 + 12 a_1 a_2 a_5$		 &	\multirow{2}{*}{$(24,24,24,24,24\,;-80)$}
\\
$(5,45)$          			   &	$+ 12 a_1 a_3 a_5 + 12 a_2 a_3 a_5 + 12 a_1 a_4 a_5 + 12 a_2 a_4 a_5 + 12 a_3 a_4 a_5$		&
\\ \hline
{\bf 6770}		&	$24 a_1 a_2 a_3 + 12 a_1 a_2 a_4 + 12 a_1 a_3 a_4 + 12 a_2 a_3 a_4 + 12 a_1 a_2 a_5$		 &	\multirow{2}{*}{$(24,24,24,24,24\,;-64)$}
\\
$(5,37)$          			   &	$+ 12 a_1 a_3 a_5 + 12 a_2 a_3 a_5 + 12 a_1 a_4 a_5 + 12 a_2 a_4 a_5 + 12 a_3 a_4 a_5$		&
\\ \hline
{\bf 6715,6788}	&	$12 a_1 a_2 a_3 + 12 a_1 a_2 a_4 + 12 a_1 a_3 a_4 + 12 a_2 a_3 a_4 + 24 a_1 a_2 a_5$		 &	\multirow{3}{*}{$(24,24,24,24,64\,;-64)$}
\\
{\bf 6836,\,6927}          			&	$+ 24 a_1 a_3 a_5 + 24 a_2 a_3 a_5 + 24 a_1 a_4 a_5 + 24 a_2 a_4 a_5 + 24 a_3 a_4 a_5$		&
\\
$(5,37)$	                    &	 $+ 24 a_1 a_5^2 + 24 a_2 a_5^2 + 24 a_3 a_5^2 + 24 a_4 a_5^2 + 16 a_5^3 $	&
\\ \hline
{\bf 6732,\,6802}	&	$12 a_1 a_2 a_3 + 12 a_1 a_2 a_4 + 12 a_1 a_3 a_4 + 12 a_2 a_3 a_4 + 12 a_1 a_2 a_5$	 &	\multirow{3}{*}{$(24,24,24,24,56\,;-64)$}
\\
{\bf 6834,\,6896}        			&	$+ 24 a_1 a_3 a_5 + 24 a_2 a_3 a_5 + 24 a_1 a_4 a_5 + 24 a_2 a_4 a_5 + 24 a_3 a_4 a_5$		&
\\
$(5,37)$ 						&	 $+ 12 a_1 a_5^2 + 12 a_2 a_5^2 + 24 a_3 a_5^2 + 24 a_4 a_5^2 + 8 a_5^3$	&
\\ \hline
{\bf 6225}	                    &	$12 a_1 a_2 a_3 + 18 a_1 a_2 a_4 + 18 a_1 a_3 a_4 + 12 a_2 a_3 a_4 + 6 a_1 a_4^2$		 &	\multirow{4}{*}{$(24,24,24,36,36\,;-56)$}
\\
         			&	$+ 6 a_2 a_4^2 + 6 a_3 a_4^2 + 18 a_1 a_2 a_5 + 18 a_1 a_3 a_5 + 12 a_2 a_3 a_5$		 &
\\
$(5,33)$		        &	 $+ 24 a_1 a_4 a_5 + 24 a_2 a_4 a_5 + 24 a_3 a_4 a_5 + 12 a_4^2 a_5 + 6 a_1 a_5^2$	     &
\\
	                    &	$+ 6 a_2 a_5^2 + 6 a_3 a_5^2 + 12 a_4 a_5^2$	                                         &
\\ \hline
{\bf 5302}	&	$12 a_1 a_2 a_3 + 12 a_1 a_2 a_4 + 12 a_1 a_3 a_4 + 12 a_2 a_3 a_4 + 12 a_1 a_2 a_5$ 	 &	\multirow{4}{*}{$(24,24,24,24,24,24\,;-48)$}
\\
 		         			&	$+ 12 a_1 a_3 a_5 + 12 a_2 a_3 a_5 + 12 a_1 a_4 a_5 + 12 a_2 a_4 a_5 + 12 a_3 a_4 a_5$		&
\\
$(6,30)$								&	 $+ 12 a_1 a_2 a_6 + 12 a_1 a_3 a_6 + 12 a_2 a_3 a_6 + 12 a_1 a_4 a_6 + 12 a_2 a_4 a_6$		&
\\
								&    $+ 12 a_3 a_4 a_6 + 12 a_1 a_5 a_6 + 12 a_2 a_5 a_6 + 12 a_3 a_5 a_6 + 12 a_4 a_5 a_6$		&
\\ \hline
 \end{tabular}
\renewcommand{\arraystretch}{1}
\end{center}
\caption{\label{tb:CICYs_summary}
This table summarizes the basic topological data for the CICY geometries with $4 \leq h_{11} \leq 5$ used for model building in this paper. The second column gives the total Calabi-Yau volume from which the intersection numbers $\kappa_{ijk}$ can be read off. The third column gives the integrated second and third Chern classes.}
\end{table}

A subset of Calabi-Yau manifolds, which is often considered in phenomenological applications, can be described as hypersurfaces (or complete intersections) in products of projective ambient spaces $\otimes_a \mathbbm{P}^{k_a}$. Such complete intersection Calabi-Yaus (CICYs) are defined by the degrees $\Gamma_{aA}$ of the polynomials, labeled by $A$, that characterize the complete intersection under the projective scalings. Since we are considering favorable descriptions of CICYs in which all CY divisors are given as pullbacks of the projective ambient space hyperplane divisors, we have $h_{11}$ projective ambient space factors. In order to end up with a CY 3-fold, we need a complete intersection of codimension $\sum k_a-3$. The data defining the CICY can thus be summarized succinctly in terms of the $h_{11}\times(\sum k_a-3)$ configuration matrix $\Gamma = (\Gamma_{aA})$ in which each row specifies one $\mathbbm{P}^{k_a}$ factor and each column represents the degrees of one polynomial equation under the various projective scalings. 

One of the simplest examples of this is the quintic $(4|5)$ which corresponds to a degree five homogeneous polynomial in $\mathbbm{P}^4$. Often, the first column which specifies the dimension of the projective space is omitted since for a CY it follows unambiguously from the sum of the degrees of the defining polynomials. In this case, the quintic has just $(5)$ as its configuration matrix.

In this paper we use CICYs from \cite{CYPage}. The data presented there is based on classifications \cite{Candelas:1987kf,Braun:2010vc,CYweb}. Out of the 7890 CICYs we focus on a subset of the 74 manifolds which are favorable and allow for free discrete actions. They all have $h_{11} \leq 6$.
For the CICYs with $h_{11} \leq 3$ we found that it is impossible to satisfy the model building conditions (Bianchi identities and DUY equations) while getting to the desired number of chiral SM families. Thus, we refrain from explicitly discussing those manifolds.

In an accompanying paper~\cite{Blaszczyk:2015zta} we have reviewed some useful formulae to obtain the relevant topological data, following the techniques described in~\cite{Hosono:1994ax}. A summary of the topological data, intersection numbers and second/third Chern classes of the CICYs studied in this paper can be found in Table~\ref{tb:CICYs_summary}. In particular, the intersection numbers $\gk_{ijk}$ can be read off from the total volume $\text{Vol}(X)$ of the corresponding CICY. For the favorable CICYs under consideration we always assume that we work in a minimal integral basis \eqref{MinIntBasis}. 

\subsection{The Schoen manifold} 
\label{sc:Schoen} 

As an alternative Calabi-Yau example we consider the so-called Schoen manifold. This manifold can be thought of as the hypersurface with configuration matrix:
\equ{
\renewcommand{\arraystretch}{1}
\left( \begin{array}{cc}
3 & 0 \\ 
0 & 3 \\ 
1 & 1 
\end{array}
\right)~.  
\renewcommand{\arraystretch}{\arrystrch}
}
In this description of the Schoen manifold all complex structure moduli are encoded explicitly as deformation parameters of the two homogeneous polynomials. However, only three of the total 19 K\"ahler moduli are realized explicitly. 

A different realization of the Schoen manifold that describes all K\"ahler parameters explicitly, is given in terms of a resolution of a particular $\Intr_2\times \Intr_2$ orbifold, the (0-2) orbifold in the Donagi-Wendland classification~\cite{Donagi:2008xy,Fischer:2012qj}. This description has the advantage that all $h_{11}=19$ divisors are described explicitly. These divisors are grouped into three types, denoted by $R_{1,2,3}$, $E_r$  and $\widetilde{E}_{r}$, where $r = (r_1r_2r_3)$ with $r_i = 0,1$. The divisors $R_i$ are often referred to as inherited divisors as they arise from torus divisors on the underlying orbifold. The divisors $E_r$  and $\widetilde{E}_{r}$ are called exceptional divisors as they correspond to the blowup cycles that appear in the resolution process. Even though the set $\{R_i, E_r, \widetilde{E}_r\}$ provides a convenient basis for many purposes, it should be stressed that it does not constitute a minimal integral basis. The intersection numbers can again be read off from the total volume of the Schoen manifold given in Table~\ref{tb:Schoen_info} using the expansion 
\equ{ 
J = \sum a_i\, R_i - \sum b_r\, E_r - \sum \widetilde b_{r} \, \widetilde E_{r}~
}
of its K\"ahler form. Here all  $b_r, \widetilde{b}_{r}$ and $a_i$ are positive and subject to the Mori cone conditions, some of which are listed in Table~\ref{tb:Schoen_info}. 
Additional conditions are mentioned there to ensure that the total volume of $X$ is positive.

\begin{table}
\begin{center}
\renewcommand{\arraystretch}{1.2}
\begin{tabular}{|p{2cm}||c|c|}
\hline
\centering{\bf }			&	{\bf Total volume}																	&	{\bf Chern classes}
\\ \hline
\multirow{3}{2cm}{\centering{\bf Schoen \\ manifold}\\$(19,19)$} 			&	\multirow{2}{*}{$6\,\text{Vol}(X) = 12\, a_1\Big\{ a_2 a_3 - \sum(b_r)^2 \Big\}$}	&	$c_{2}(R_1)=c_{2}(R_2) =24$, $c_{2}(R_3)=0$
\\
		&   																				&	$c_{2}(E_r)=c_{2}(\widetilde{E}_{r})=0$
\\
		&	\;\qquad\qquad$+  12\, a_2\Big\{ a_1 a_3 - \sum (\widetilde{b}_{r})^2 \Big\}$						&	$c_3=0$
\\ 
\hline\hline
\multicolumn{3}{|c|}{\bf Some necessary conditions on the K\"ahler parameters}
\\ \hline  
$\text{Vol}(X)>0$	&	$\sum (b_r)^2 < a_2a_3$ 	&	$\sum (\widetilde{b}_r)^2 < a_1a_3$
\\ \hline
			&	$\sum_{r_1,r_2} b_{r_1r_2r_3} < a_2$	&	$\sum_{r_1,r_2} \widetilde{b}_{r_1r_2r_3} < a_1$
\\
$\text{Vol}(C)>0$	&	$\sum_{r_3} b_{r_1r_2r_3} < a_3$ 		&	$\sum_{r_3} \widetilde{b}_{r_1r_2r_3} < a_3$
\\
			&	$\sum_{r_3} b_{r_1r_2r_3} + \sum_{r_3} \widetilde{b}_{r_1r_2r_3} < a_3$	&	$a_i,b_r,\widetilde{b}_r>0$
\\ \hline
\multirow{2}{*}{$\text{Vol}(D)>0$}			&	$\sum_{r_1,r_2}  \text{Vol}(E_{r_1r_2r_3}) <  \text{Vol}(R_3)$	&	 $\sum_{r_1,r_2}  \text{Vol}(\widetilde{E}_{r_1r_2r_3}) <  \text{Vol}(R_3)$
\\
	&	$\sum  \text{Vol}(E_r)^2 < 2\,  \text{Vol}(R_1)  \text{Vol}(R_3)
$		&	$\sum  \text{Vol}(\widetilde{E}_r)^2 < 2\,  \text{Vol}(R_2)  \text{Vol}(R_3)
$
\\ \hline 
\end{tabular}
\renewcommand{\arraystretch}{1}
\end{center}
\caption{\label{tb:Schoen_info}
This table both characterizes the basic topological data of the Schoen manifold and gives a number of necessary conditions to be inside the K\"ahler and Mori cones. 
}
\end{table}

\section{Heterotic line bundle models}
\label{sc:HetLBmodels} 

\subsection{Line bundle vectors}

Next, we briefly characterize heterotic models on smooth compactifications of any of the three heterotic strings with line bundles. Concretely, we consider the gauge background
\equ{ \label{LineBundleFlux} 
 \frac{\mathcal{F}}{2\pi} = \widehat D_i\, H_i~, 
 \qquad 
 H_i = V_i^I\, H_I
 }
embedded in the Cartan subalgebra of the ten-dimensional gauge group $\mathcal{G}$. 
This gauge flux is characterized by a set of bundle vectors $V_i =(V_i^I)$, one for each divisor (1,1)-form $\widehat D_i$ (see e.g.\ \cite{Nibbelink:2007rd,Nibbelink:2009sp}). For the E$_8\times$E$_8$ and SO(16)$\times$SO(16) theories these bundle vectors can be conveniently decomposed into two pieces $V=(V_i^{\prime}, V_i^{\prime\prime})$ corresponding to the ten-dimensional gauge group factors.

To ensure integral multiplicities for all massless states the bundle vectors are subject to the flux quantization conditions,
\equ{ \label{FluxQuantization} 
q_i(C)\, V_i \in \gL~, 
}
for any curve $C$ where $q_i(C)$ is defined in~\eqref{CurveIntersections}. The possible lattices $\gL$ are given by
\begin{equation}\label{GaugVectors_on_lattice}
 \gL_\text{E$_8\times$E$_8$} =  (\mathbf{R}_8 \oplus \mathbf{S}_8) \otimes (\mathbf{R}_8\oplus \mathbf{S}_8)~, 
\quad 
\gL_\text{SO(32)} = \mathbf{R}_{16} \oplus \mathbf{S}_{16}~, 
\quad 
 \gL_\text{SO(16)$\times$SO(16)} =  
 \big(\mathbf{R}_8 \otimes \mathbf{R}_8\big) \oplus \big(\mathbf{S}_8 \otimes \mathbf{S}_8\big)
\end{equation}
for the E$_8\times$E$_8$, SO(32) or SO(16)$\times$SO(16) heterotic string, respectively. If $\{D_k\}$ defines a minimal integral basis of divisors~\eqref{MinIntBasis}, like for the CICY geometries under inspection, the line bundle vectors lie automatically inside these lattices. For the Schoen manifold we do not employ a minimal integral basis, hence \eqref{FluxQuantization} results in more complicated conditions on the line bundle input data, see~\cite{Nibbelink:2012de} for a detailed exposition of these conditions. 
 
\subsection{Gram matrix}

It turns out to be worthwhile to define the Gram matrix associated to a set of line bundle vectors as 
\equ{ \label{GramMatrix} 
K_{ij} = K_{ji} = V_i \cdot V_j=\sum_I V_i^I V_j^I~, 
} 
with standard Euclidean inner product. Consequently, we have 
\equ{\label{K_matrix_conditions}
0 \leq K_{ii}~,
\qquad 
| K_{ij}|^2 \leq K_{ii}\, K_{jj}~, 
}
for $i,j = 1,\ldots, h_{11}$. The first condition is the statement that the vector-norm $V_i^2$ is non-negative. The second condition is a rewriting of the Schwarz inequality $\vert V_i \cdot V_j \vert^2 \leq \vert V_i \vert^2 \vert V_j \vert^2$.

\begin{table}
\begin{center}
\renewcommand{\arraystretch}{1.2}
\begin{tabular}{|c|c|c|}
\hline
\multicolumn{3}{|c|}{\bf Model-building constraints on input data}
\\ 
\multirow{2}{*}{\bf Constraint}	&	{\bf Condition on bundle vectors}			&	{\bf Condition on Gram matrices}
\\
							&	$V_i$\, with $i = 1,\ldots, h_{11}$	&   $K_{ij}=K_{ji}$\, with  $i,j = 1,\ldots, h_{11}$
\\ \hline\hline
\multirow{2}{*}{\bf Flux quantization} 	&	\multirow{2}{*}{$V_i$ even lattice vector}		&	$\vert K_{ij} \vert^2 \leq K_{ii}\, K_{jj}$\,,	
\\
									&	
									&	$K_{ij} \in \mathbb{Z}$ and $K_{ii} \in 2\, \mathbb{Z}_{0}^{+}$
\\ \hline
\multirow{2}{*}{\bf Bianchi identities} 	&	\multirow{2}{*}{$\gk_{ijk} \, V_j \cdot V_k + 2\, c_{2i} = 0$	}&	\multirow{2}{*}{$\gk_{ijk} \, K_{jk} + 2\, c_{2i} = 0$} 
\\&& 
\\ \hline 
\multirow{2}{*}{\bf Tree-level DUY} 	 	&	\multirow{2}{*}{$\dsp \sum_i \text{Vol}(D_i)\, V_i^I = 0$}	&	\multirow{2}{*}{$\dsp K_{ii} = -\frac1{\text{Vol}(D_i)} \sum_{j \neq i} \text{Vol}(D_j)\, K_{ij}$} 
\\ && 
\\ \hline
\end{tabular}
\renewcommand{\arraystretch}{1}
\end{center}
\caption{\label{tb:ModelBuildingConstraints}
The table summarizes the main model-building constraints. In the second column the conditions are stated in terms of the line bundle vectors, while in the third column we have rewritten the conditions in terms of the $K$-matrix, which is used for generating random models.
}
\end{table}

All well-known conditions on a line bundle background can be written in terms of conditions on the associated Gram matrix. This is summarized in Table~\ref{tb:ModelBuildingConstraints} for the cases where we have an integral basis~\eqref{MinIntBasis}. In detail we have:

\subsubsection*{Flux quantization}

The flux quantization conditions~\eqref{FluxQuantization} imply that the entries of the Gram matrix are restricted such that 
\equ{ \label{GramQuantization}
K_{ij}\, q_j(C) \in \Intr \quad \forall i~, 
}
for any curve $C$, since all three lattices~\eqref{GaugVectors_on_lattice} are integral. Furthermore $q_j(C)\in\Intr$, since they are given in terms of intersection numbers, cf.~\eqref{CurveIntersections}. Thus, we always fulfill these conditions by taking integer entries for the Gram matrix. In the minimal integral basis~\eqref{MinIntBasis} the entries of $K$ necessarily satisfy  
\equ{ 
K_{ij} \in \Intr~, 
\qquad 
K_{ii} \in 2\, \Intr_{\geq 0}~. 
}
The second condition arises because all three possible lattices~\eqref{GaugVectors_on_lattice} are even. 
 
\subsubsection*{Bianchi identities}

We assume throughout this work that we do not have any NS5-branes. In this case, the Bianchi identities for the $B$-field can be represented as 
\equ{ \label{Kmatrix} 
\gk_{ijk} \, K_{jk} + 2\, c_{2i} = 0~, 
}
in terms of the Gram matrix~\eqref{GramMatrix}. 

\subsubsection*{Donaldson-Uhlenbeck-Yau equations}

Finally, the gauge background has to satisfy the DUY equations in order to guarantee that a solution to the underlying Hermitian Yang-Mills equations can be found. The tree-level DUY equations without additional VEVs read
\equ{\label{DUY1}  
 V_j^I\, \text{Vol}(D_j)  = 0~,
}
for all $I$ where the divisor volumes Vol$(D_i)$, defined in \eqref{VolumesWithDilaton}, will be always taken deep inside the K\"ahler cone to ensure that our geometric description makes sense and that the supergravity approximation is valid. 
Dotting \eqref{DUY1} with $V_i$ for every $i=1,...,h_{11}$ we obtain a necessary and sufficient condition for the DUY equations in terms of the Gram matrix,
\equ{\label{DUY2}  
 K_{ij}\, \text{Vol}(D_j)
  = 0~. 
}

In this paper we do not consider the one-loop correction to DUY equations. 
The main reason for this is that for the non-supersymmetric SO(16)$\times$SO(16) theory it is unclear what the order one-loop corrections to DUY are. 
Since with the present work we aim to compare the three heterotic theories side-by-side, we also refrain from including the loop corrections for the supersymmetric heterotic theories as well. 

\subsection{Equivariant line bundles and Wilson lines}\label{sc:EquivariantBundles}

By construction, CICYs have a trivial fundamental group. It can become non-simply connected by modding out some compatible freely acting symmetry (which does not introduce any fixed points). In this paper we focus on $\mathbb{Z}_N$ symmetries, i.e. one-generator Abelian symmetries of order $N$.

Generically, such a symmetry consists of introducing phases and permuting homogeneous coordinates inside the ambient $\mathbb{P}^N$ factors. It could also permute $\mathbb{P}^N$ factors, but will never mix the coordinates among different $\mathbb{P}^N$ factors. In the ambient space divisor basis $\lbrace D_i \rbrace$, the first two actions do not pose any additional constraint.
In contrast, if two $\mathbb{P}^N$ factors are permuted, the corresponding divisors are also interchanged.
In the latter case, we have to ensure that our line bundle background, constructed on the initial basis $\lbrace D_i \rbrace$ (in the so-called upstairs picture) is still a valid gauge background after the discrete symmetry is modded out (in the so-called downstairs picture).
This compatibility of the line bundle background with the symmetry action of the underlying geometry is called equivariance. 

We restrict ourselves to modding out discrete symmetries of the lowest order possible: For the subset we consider in Table~\ref{tb:CICYs_summary} these are $\mathbb{Z}_2$ actions. In all but one case the equivariance constraint is trivial, i.e.\ no new condition arises for our bundle vectors $V_i$\,. 
Only for CICY~6225 we have to impose $V_4=V_5$ to ensure equivariance of the line bundle background, since the $\mathbb{Z}_2$ action maps $D_4 \leftrightarrow D_5$. The freely acting $\mathbb{Z}_2$ symmetry admitted by the Schoen manifold also imposes non-trivial conditions on the 19 bundle vectors, see Ref.~\cite{Nibbelink:2012de} for details.

\section{Automated construction of smooth compactification models}
\label{sc:RandomModels} 

In this section we describe our procedure to perform model scans for smooth Calabi-Yau compactifications with line bundle gauge fluxes of the three heterotic theories. The phenomenological explorations of the supersymmetric SO(32) theory and  the non-supersymmetric SO(16)$\times$SO(16) string on smooth Calabi-Yau geometries have not been considered in the past.  The E$_8\times$E$_8$ string models are rather well established, therefore, we use E$_8\times$E$_8$ model building to cross check our methods and computer codes. The final aims of our studies are the following: 
\begin{itemize}
\item
Generate four-dimensional models on smooth geometries for any of the three heterotic strings in an automated way. 
\item 
Search for MSSM-like models for the E$_8\times$E$_8$ and SO(32) theories. 
\item 
Search for tachyon-free SM-like models for the SO(16)$\times$SO(16) theory. 
\item 
Establish estimates of how fruitful and phenomenologically relevant each theory might be on a given geometry. 
\end{itemize} 
In order to perform such investigations in an automated fashion we essentially go through three stages: 
\begin{enumerate} 
\item {\bf Generation of the model input data} 
\\[1ex]
Starting from a given heterotic theory and geometry we generate a large collection of compatible bundle vectors. 
\item {\bf Computation of the resulting spectrum}
\\[1ex]
For each of the corresponding models we compute the full charged massless chiral spectrum and check whether it is free of anomalies. 
\item {\bf Analysis of the phenomenology}
\\[1ex] 
Based on these spectra we investigate some of their elementary phenomenological properties, e.g.\ to what extend the massless spectra can be related to (MS)SM physics.
\end{enumerate}
In the following subsections we describe each of these steps in more detail.

\subsection{Generation of the model input data}

We focus on line bundle backgrounds on smooth Calabi-Yau manifolds as discussed in Section~\ref{sc:HetLBmodels}. To generate models we make use of the description spelled out there: Instead of directly attempting to find a set of line bundle vectors that satisfy the Bianchi identities and then check whether they fulfill the tree-level DUY equations, we mostly try to formulate the problem on the level of the Gram matrix $K$ defined in \eqref{GramMatrix}. 

\begin{figure}[t!]
\begin{center} 
\includegraphics[width=0.8\textwidth]{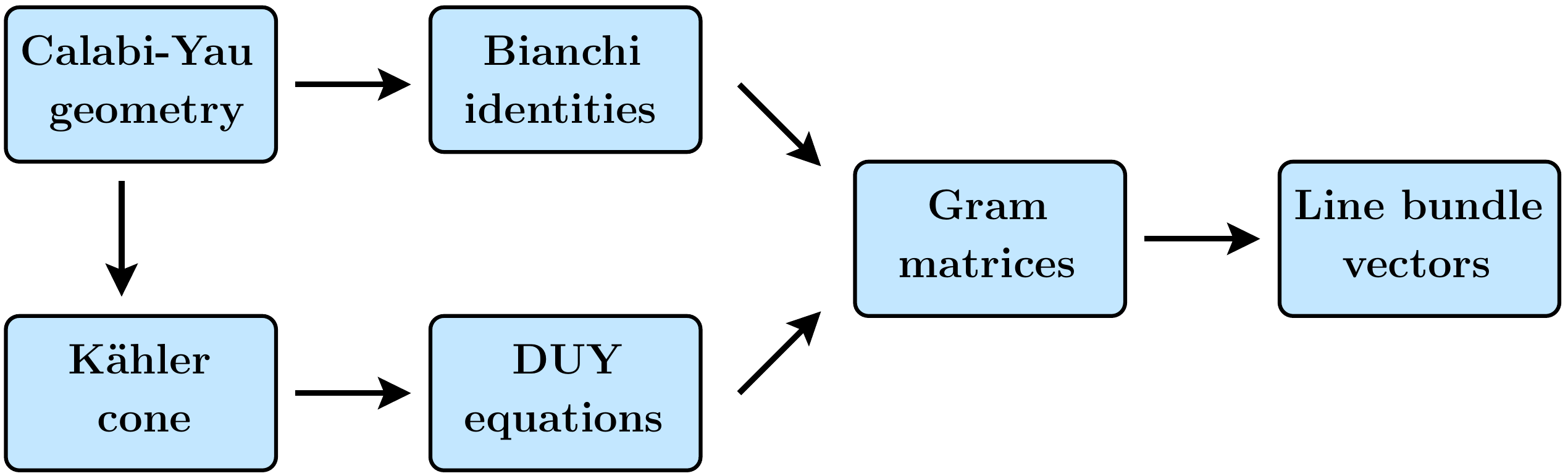}
\end{center}
\caption{\label{fg:ModelGeneration}
This diagram displays the steps involved in going from an input geometry to a full specification of the upstairs line bundle model. 
} 
\end{figure}

For a given $K$ we can construct a set of bundle vectors that lead to this Gram matrix, which then fully define an upstairs model, i.e.\ before modding out any freely acting symmetries. The procedure is summarized in the diagram in Figure~\ref{fg:ModelGeneration}. Even though this diagram indicates the logical order of the various construction steps, the use of the Gram matrix means that some steps can be performed independently of each other and therefore in parallel. In particular, at the level of $K$, the consistency conditions displayed in Table~\ref{tb:ModelBuildingConstraints} are identical for all three heterotic strings. Therefore, finding an admissible set of Gram matrices is a theory-independent process. Before we discuss the various steps in more detail, we should also stress that Figure~\ref{fg:ModelGeneration} indicates the generic steps involved. For some geometries certain steps can be optimized or even by-passed. We will discuss the cases where we have done so below.

\subsubsection*{Generating the K\"ahler cone}\label{sc:KahlerConeGeneration}

Since we do not know how to efficiently solve the divisor volume constraints, 
we simply construct the K\"ahler cone for the manifolds under investigation explicitly. More precisely, we generate a large number of points inside the K\"ahler cone.  For the geometries where we have the explicit parametrization of the Mori cone in a minimal basis as $a_i > 0$, we do so by running through a finite integer range for the K\"ahler parameters and normalizing the resulting $h_{11}$ volumes Vol$(D_i)$ by their greatest common divisor.

\subsubsection*{Generating Gram matrices}\label{sc:GramMatrixGeneration}

Solving the Bianchi identities and DUY equations is essentially immediate once one has specified the geometry and a point in the K\"ahler cone. Hence, we can turn to the construction of the Gram matrices. The Gram matrix has $h_{11}(h_{11}+1)/2$ independent entries. Generically, the Bianchi identities~\eqref{Kmatrix} and the DUY equations~\eqref{DUY2} fix $2\, h_{11}$ entries, such that we have $h_{11}(h_{11}-3)/2$ independent components in general.

For many geometries under consideration, the Bianchi identities only depend on a small number of linear combinations of the diagonal entries $K_{ii}$ of the Gram matrix. Hence, in these cases it is beneficial to solve mainly for these diagonal entries in terms of the off-diagonal entries $K_{ij}$, $i\neq j$, using the DUY equations~\eqref{DUY2}: 
\equ{\label{DUY_sol}
K_{ii} = -\frac 1{\text{Vol}(D_i)} \sum_{j \neq i} \text{Vol}(D_j) K_{ij}~. 
}
Since we have generated the K\"ahler parameters $a_i$ such that we are inside the K\"ahler cone, dividing by any volume $\text{Vol}(D_i)$ never poses a problem.  

In relatively rare cases the Bianchi identities and the DUY equations are partially linear dependent. This happens when the combined equations do not have maximal rank $2\, h_{11}$. In this case the Bianchi identities dictate some combinations of the divisor volumes, which could lead to solutions outside of the K\"ahler cone and consequently outside of the validity of our approximation.

As emphasized in~\cite{Nibbelink:2015ixa}, there seems not to exist a clear bound for the range of the entries $K_{ij}$ left undetermined by the simultaneous solution of the Bianchi identities and the DUY equations. This range needs for sure to be at least such that the second Chern class contribution with mostly $c_{2i} \geq 0$ can be compensated. In practice, the possible values of $K$-matrix elements have to be taken from a finite range. To be able to solve the combined system efficiently and fast, we choose this range according to the number of free parameters that we have to scan over while inspecting the Bianchi identities. The more adjustable $K_{ij}$ entries there are, the smaller the range practically has to become. 

\subsubsection*{Generating bundle vectors}
\label{sc:BundleVectorsGeneration}

For a given Gram matrix $K$, we can always construct a collection of line bundle vectors $\lbrace V_i \rbrace$ on the appropriate lattice. For this it is crucial that $K$ obeys the conditions~\eqref{K_matrix_conditions}: Most randomly generated matrices cannot be written as inner products of vectors as given in~\eqref{GramMatrix}. 
The $K$-matrix construction ensures that a set of linearly depended line bundle vectors $\lbrace V_i \rbrace$ always exists such that the integrated Bianchi identities and the tree-level DUY equations are solved inside the K\"ahler cone. 
Once we found a set of admissible $K$-matrices, we make use of essentially two algorithms to determine a generating set of line bundle vectors $\lbrace V_i \rbrace$ on the appropriate lattice: 

The first approach starts from a vector $V^0_i$ ($i=1,...,h_{11}$) and tries to find lattice directions which minimize $f \equiv \sum_{j=1}^{j<i} \vert V_i \cdot V_j - K_{ij} \vert$\,. This is done by adding or subtracting simple roots from $V_i^0$ until a preferred directions has been singled out. Since we are working on a lattice, we expect that at some point $f=0$\,. Although this algorithm is extremely fast and efficient for small values of $K_{ij}$\,, it quickly becomes rather slow for larger norms (e.g.\ for $K_{ii} > 10$). 

For the second approach, we randomly pick vectors of given norm $K_{ii}$ and try to fix the off-diagonal scalar products $K_{ij}$ through an adjustable trial-and-error procedure. Concretely, we generate two vectors $V_1$ and $V_2$ with $V_1^2 = K_{11}$ and $V_2^2 = K_{22}$. Then we check whether $V_2 \cdot V_1 = K_{21}$\,. If this is not the case, a new $V_2^2 = K_{22}$ is generated and re-checked until the matching scalar product $V_1 \cdot V_2$ is found. We proceed analogously for the other bundle vectors. A potential deficiency of this algorithm is that it might fail to converge: When $i$ vectors have been successfully constructed, it is not guaranteed that a vector $V_{i+1}$ with $V_{i+1}^2=K_{i+1,i+1}$ exists that produces the corresponding $K$-matrix elements $K_{i+1, j}$ with $j=1,\ldots,i$\,. Therefore, the algorithm tries to statistically establish (based on time optimization criteria) the maximum number of randomly generated vectors, before the whole ansatz (including the $i$ successfully constructed vectors) is abandoned.  Once  $h_{11}-1$ vectors have been successfully determined, the final one is computed via the tree-level DUY equations using~\eqref{DUY_sol}. 

In practice, we need to generate line bundle vectors for a given $K$-matrix in a reasonable amount of time (i.e.\ less than $\sim 1s$).
Hence, we are obliged to keep the $K_{ij}$'s sufficiently small. Testing both approaches, we find that the second procedure is significantly faster in generating admissible line bundle backgrounds, allowing for bigger values of $K$-matrix elements as well as higher $h_{11}$ ($K_{ij} \leq 20$ and $h_{11} \leq 6$ on favorable CICYs) without significant loss of efficiency.

\subsection{Computation of the four-dimensional spectrum}

\subsubsection*{Unbroken gauge group}

The unbroken gauge group in four dimensions is determined by the subgroup of the 10D gauge group that commutes with the line bundle background~\eqref{LineBundleFlux}. Denoting the roots of the ten-dimensional gauge group by $\ga$, the associated unbroken generators in four dimensions are determined by
\equ{\label{GGBreaking}
\ga \cdot V_i = 0 \qquad\forall i=1,\ldots,h_{11}~.
}

Furthermore, as discussed in Section~\ref{sc:EquivariantBundles}, on non-simply-connected CYs we can turn on non-trivial gauge backgrounds with zero field strength, i.e.\ Wilson lines. We focus on freely acting  $\mathbb{Z}_N$ symmetries of  the underlying geometry. Then, the corresponding Wilson line $W$ has to be of the same order, $N W \in \gL$ with $\gL$ defined in \eqref{GaugVectors_on_lattice}. In addition to the projection conditions \eqref{GGBreaking} due to gauge fluxes, the generators of the surviving group now have to be invariant under the action of the freely acting Wilson line $W$,
\equ{
\ga \cdot W = 0~~\text{mod } 1\,.
}

\subsubsection*{Charged chiral spectrum}

\begin{table}[t!]
\begin{center}
\renewcommand{\arraystretch}{1.8}
\begin{tabular}{|c||c|c|c|}
\hline
\multicolumn{4}{|c|}{\bf Weights of the 10D massless charged heterotic states: $(p^2 = 2)$}  \\
\hline
{\bf State}	&	{\bf N=1, E$\boldsymbol{_8\times}$E$\boldsymbol{_8}$}	& {\bf N=1, SO(32)}	   & 	{\bf N=0, SO(16)$\boldsymbol{\times}$SO(16)}	
\\ \hline\hline 
\cellcolor{lightgray} 
{\bf Gauge}	
&\cellcolor{lightgray}	
$\big(\underline{\pm 1,\pm 1, 0^6}\big)\big(0^8\big)$\,, $\big(\underline{\sm \sfrac12^{2k},\sfrac 12^{8-2k}}\big)\big(0^8\big)$	
&\cellcolor{lightgray}	
$(\underline{\pm 1, \pm 1, 0^{14}})$		
&\cellcolor{lightgray}	
$(\underline{\pm 1,\pm 1, 0^6})(0^8)$ 
\\[-.2ex] 
\cellcolor{lightgray}	
{\bf bosons}							
&\cellcolor{lightgray}	
$\big(0^8\big)\big(\underline{\pm 1,\pm 1, 0^6}\big)$\,, $\big(0^8\big)\big(\underline{\sm \sfrac12^{2k},\sfrac 12^{8-2k}}\big)$	
&\cellcolor{lightgray}		
&\cellcolor{lightgray}	
$(0^8)(\underline{\pm 1,\pm 1, 0^6})$ 
\\\hline\hline  
{\bf Pos.\ chiral}	      	
&	
$\big(\underline{\pm 1,\pm1, 0^6}\big)\big(0^8\big)$\,, $\big(\underline{\sm \sfrac12^{2k},\sfrac 12^{8-2k}}\big)\big(0^8\big)$		
&					
&	
$\big(\underline{\sm \sfrac12^{2k},\sfrac 12^{8-2k}}\big)\big(0^8\big)$
\\[-.2ex] 
{\bf fermions}    			
&	
$\big(0^8\big)\big(\underline{\pm 1,\pm 1, 0^6}\big)$\,, $\big(0^8\big)\big(\underline{\sm \sfrac12^{2k},\sfrac 12^{8-2k}}\big)$		
&					
&								
$\big(0^8\big)\big(\underline{\sm \sfrac12^{2k},\sfrac 12^{8-2k}}\big)$
\\ \hline 
{\bf Neg.\ chiral}  
&			
&
\multirow{2}{*}{$\big(\underline{\pm 1, \pm 1, 0^{14}}\big)$}						
&	
\multirow{2}{*}{$\big(\underline{\pm 1, 0^7}\big)\big(\underline{\pm 1, 0^7}\big)$} 
\\[-.2ex] 
{\bf fermions}  
&
&
&
\\
\hline
\end{tabular}
\renewcommand{\arraystretch}{1.0}
\end{center}
\caption{\label{tb:Lattices}
The weights of the charged massless ten-dimensional states of the three heterotic theories are listed. 
Underlined entries means permutation of the corresponding entries. A power of an entry means repetition of this entry and $0 \leq k \leq 4$. 
We have used the convention that the chirality of fermions of the supersymmetric E$_8\times$E$_8$ and SO(32) theories are opposite; the non-supersymmetric SO(16)$\times$SO(16) theory contains fermions of both chiralities. 
}
\end{table}

Subsequently, the charged massless spectrum in four dimensions is computed using the multiplicity operator. The charged massless states in ten dimensions are characterized by certain sets $R$ of lattice vectors $p\in \gL$ and the ten-dimensional chirality ``spin'' $s$ for fermions (for bosons we simply take $s=1$). This is in particular important for the non-supersymmetric SO(16)$\times$SO(16) theory which contains fermionic states of both chiralities. We take the chirality of the gauginos of the E$_8\times$E$_8$ theory positive and those of the SO(32) theory negative. The relevant weights $p$ (with $p^2=2$) of the various representations for the three heterotic theories are indicated in Table~\ref{tb:Lattices}. 

The multiplicity operator evaluated on such states takes the form
\equ{ \label{MultiOp4D}
 \mathcal{N} =  \mathcal{N}(p;s)  
  = \frac s6\, \kappa_{ijk}\, (p\cdot V_i) (p\cdot V_j) (p\cdot V_k) + \frac s{12} c_{2i}\,  (p\cdot V_i)~. 
}
For details on how this formula is obtained, see e.g.~\cite{Nibbelink:2007rd}. Since it counts the number of chiral states we can take those $p$ for which $\mathcal{N}\geq0$ without loss of generality; states with $\mathcal{N}<0$ then corresponds to the CPT conjugates of the former. In~\cite{Blaszczyk:2014qoa,Blaszczyk:2015zta} it is explained why this formula can also be used in the non-supersymmetric context to compute the massless spectra of both bosons and fermions provided that the geometrical background itself is Calabi-Yau. 
Finally, to obtain the chiral spectrum in the downstairs picture after modding out a freely acting $\mathbb{Z}_N$ symmetry,  we simply need to divide all upstairs multiplicities by the symmetry order $N$.

In order to ensure that our construction is consistent, the chiral spectrum is under the scrutiny of the following consistency checks: 

\subsubsection*{Integral multiplicities}

The multiplicity operator should always return integral values evaluated on all states. Given that factors of $1/6$ and $1/12$ occur in~\eqref{MultiOp4D}, this provides a non-trivial cross-check on whether the flux quantization conditions~\eqref{FluxQuantization} have been implemented properly. Moreover, considering a geometry that admits a freely acting symmetry and choosing an equivariant line bundle (as described in section~\ref{sc:EquivariantBundles}), the multiplicities should still be integral after modding out the freely acting Wilson line.

\subsubsection*{Anomaly cancellation}

\begin{figure}[t!]
\begin{center} 
\includegraphics[width=.6\textwidth]{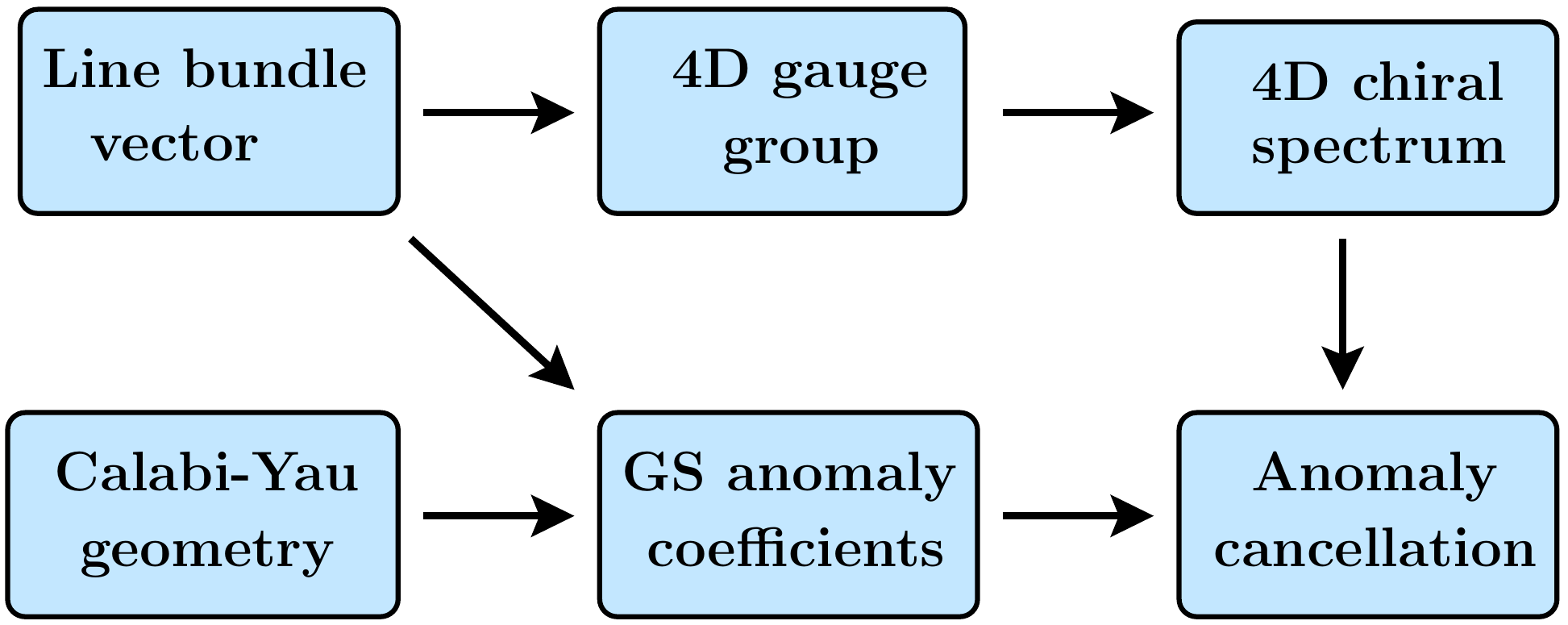}
\end{center}
\caption{\label{fg:Spectrum}
This diagram displays the information entering the anomaly checks. 
} 
\end{figure}

The second consistency check is the absence of anomalies induced by the chiral fermionic spectrum in four dimensions. These anomaly checks involve the generalized Green-Schwarz mechanisms as worked out in~\cite{Blaszczyk:2015zta} for all three heterotic theories (inspired by the works of~
\cite{Blumenhagen:2005ga,Blumenhagen:2006ux}). The information that enters the Green-Schwarz anomaly cancellation is schematically indicated in Figure~\ref{fg:Spectrum} at the level of our input data and the resulting model construction.
Consider the situation with an unbroken gauge group with $N_G$ simple non-Abelian gauge group factors and $N_U$ U(1)-factors,
\equ{ \label{UnbrokenG}
\prod_{x=1}^{N_G}G_x \times U(1)^{N_U}~, 
\quad\text{with}\quad 
\sum_x\text{rk}(G_x) + N_U = 16~.
}
Since the total rank of the gauge group is sixteen in the heterotic theories, the second relation in equation~\eqref{UnbrokenG} follows, where $\text{rk}(G_x)$ denotes the rank of the $x$-th simple group factor. 
Anomaly cancellation then leads to 
\equ{
\#(\text{checks}) = 
N_U + 
N_G + 
N_U\, N_G +
N_U + 
N_U (N_U-1) + 
\frac 16\, N_U(N_U-1)(N_U-2) 
}
independent consistency conditions. The different terms count the number of different anomaly conditions 
\items{
\item $G_x$--$G_x$--$G_x$
\item grav--grav--U(1)$^I$ 
\item $G_x$--$G_x$--U(1)$^I$
\item U(1)$^I$--U(1)$^I$--U(1)$^I$
\item U(1)$^I$--U(1)$^J$--U(1)$^J$, $I \neq J$
\item U(1)$^I$--U(1)$^J$--U(1)$^K$, $I \neq J \neq K \neq I$ 
}
which are all checked in our automated searches. (Since the index $I$ runs over the full Cartan, i.e.\ from 1 to 16, and not over the Abelian factors only, there is some redundancy in the anomaly checks we have implemented.) The absence of pure non-Abelian anomalies is checked purely at the level of the four-dimensional chiral spectrum, without the need to explicitly provide any additional information on the geometry or the heterotic theory under consideration.
On the other hand, the precise expressions for the pure and mixed Abelian anomalies involve the generalized Green-Schwarz mechanism and are therefore theory-depended and require explicit details about the compactification geometry and the gauge background.

The set of conditions for the three heterotic theories, the supersymmetric E$_8\times$E$_8$, SO(32) and the non-supersymmetric SO(16)$\times$SO(16) are explicitly stated in Tables~\ref{tb:E8xE8_GSmechanism}, \ref{tb:Spin32_GSmechanism} and~\ref{tb:nonSUSY_GSmechanism}, respectively. For the pure non-Abelian anomalies the conditions listed in these tables only correspond to the anomaly cancellation in the Cartan parts of the simple gauge group factors. We check, however, the full non-Abelian anomaly cancellation in each of these factors, as stated previously. For the E$_8\times$E$_8$ and SO(16)$\times$SO(16) theories we have to make a distinction between the cases where different U(1)s reside in the same or different ten-dimensional gauge group factors. As we have stated above throughout this paper we assume that no (anti-)NS5 branes are present, which simplifies the expressions in these tables.

In Tables~\ref{tb:E8xE8_GSmechanism}--\ref{tb:nonSUSY_GSmechanism} the following notation for an irreducible representation 
$\rep{R}=(\rep{R}_1,\ldots,\rep{R}_{N_G})$ 
of the unbroken gauge group~\eqref{UnbrokenG} has been introduced: $\rep{R}_x$, $x=1,\ldots, N_G$, denotes an irreducible representation of dimension $\text{dim}\,\rep{R}_x$ under the simple non-Abelian  group factor $G_x$.
The index $\ell(\rep{R}_x)$ of the irreducible representation $\rep{R}_x$ is given by
\equ{
\ell(\rep{R}_x) = \frac{\text{dim} \rep{R}_x}{\text{dim} \rep{Ad}_x}\, C(\rep{R}_x)~, 
}
where $C(\rep{R}_x)$ is the quadratic Casimir of the algebra $\rep{Ad}_x$ associated to simple gauge group factor $G_x$ evaluated on the representation. In particular on the fundamental (vector) representations $\rep{N}$ and $\rep{2N}$ of SU($N$) and SO($2N$)-groups we have $\ell(\rep{N}) =1$ and $\ell(\rep{2N}) =2$, respectively. In addition, we define the multiplicity 
\equ{ 
n_x(\rep{R}) = \mathcal{N}(\rep{R})\, \prod_{y \neq x} \text{dim}\,\rep{R}_y~,
}
w.r.t.\ a non-Abelian factor $G_x$ of the irreducible representation $\rep{R}$, where $\mathcal{N}(\rep{R})$ denotes the value of the multiplicity operator~\eqref{MultiOp4D} evaluated on any of the weights of $\rep{R}$. 
(For further details see e.g.~\cite{Slansky:1981yr} and Appendix~A of~\cite{Blaszczyk:2015zta}.) To describe the $G_x \times G_x \times U(1)^{N_U}$ anomaly we find it convenient to choose a particular U(1) basis given by vectors $t_a^I$, with $a=1,\ldots, N_U$ and define the charges $q_\rep{R}^a = t_a \cdot p_\rep{R}$ where $p_\rep{R}$ is a representative of the weights describing the representation $\rep{R}$. On the right-hand-side of the anomalies in Tables~\ref{tb:Spin32_GSmechanism} and~\ref{tb:nonSUSY_GSmechanism} we also need the branching of the fundamental of SO(32) or SO(16). We denote the resulting fundamental representations of the non-Abelian group factors $G_x$ by $\rep{r}_x$ and use $w \in \rep{r}_x$ to refer to their weights. 

As observed in~\cite{Blaszczyk:2015zta}, the non-supersymmetric SO(16)$\times$SO(16) theory is in some sense in between the supersymmetric E$_8\times$E$_8$ and SO(32) theories. In particular, if one takes the ten-dimensional chiralities of these two supersymmetric theories opposite to each other, the addition of their $X_8$ factors appearing in the factorized anomaly polynomial $I_{12}=X_4 X_8$ precisely coincides with the $X_8$ factors of the non-supersymmetric SO(16)$\times$SO(16) theory. Indeed, using Tables~\ref{tb:E8xE8_GSmechanism}--\ref{tb:nonSUSY_GSmechanism}, it can be readily confirmed that for the SO$(16)\times$SO$(16)$ theory, the Green-Schwarz-terms on the right hand side of the equality are obtained by subtracting the corresponding contributions of the SO(32) theory from those of the  E$_8\times$E$_8$ theory. This verifies the consistent dimensional reduction of the various anomaly polynomial factors $X_8$\,.

\newpage 

\begin{table}[!ht]
\begin{center} 
\scalebox{.95}{
\(
\renewcommand{\arraystretch}{1.1}
\begin{array}{|l||rcl|}
\hline 
\multicolumn{4}{|c|}{\text{\bf Supersymmetric E$_8\times$E$_8$ theory}}
\\ \hline\hline
\multirow{2}{*}{$\text{grav}-\text{grav}-\text{U(1)}^{I'}$} & \multirow{2}{*}{$\sum_{p}\, \mathcal{N}\, p^{I'}$}	 & \multirow{2}{*}{$=$} &	 \multirow{2}{*}{$V_i^{I'} \left( 3 \kappa_{ijk} (V_j^\prime \cdot V_k^\prime) + \sfrac52 c_{2i} \right) $} 
\\ &&&
\\ \hline 
\multirow{2}{*}{$G_{x'}-G_{x'}-\text{U(1)}^{a'}$} & \multirow{2}{*}{~~~$\sum_\rep{R}\, n_{x'}(\rep{R})\, \ell (\rep{R}_{x'})\, q^{a'}_R$}   & \multirow{2}{*}{$=$} & \multirow{2}{*}{$\sfrac12\, t^{a'} \cdot V_i \left( \kappa_{ijk} (V_j^\prime \cdot V_k^\prime) + c_{2i} \right)$} 
\\ &&&
\\ \hline
\multirow{2}{*}{$G_{x'}-G_{x'}-\text{U(1)}^{a''}$} & \multirow{2}{*}{$\sum_\rep{R}\, n_{x'}(\rep{R})\, \ell (\rep{R}_{x'})\, q^{a''}_R$}   & \multirow{2}{*}{$=$} & \multirow{2}{*}{$0$} 
\\ &&&
\\ \hline  
\multirow{2}{*}{$\text{U}(1)^{I'} - \text{U}(1)^{I'} - \text{U}(1)^{I'}$} & \multirow{2}{*}{$\sum_{p}\, \mathcal{N}\, p^{I'}p^{I'}p^{I'}$} 		& \multirow{2}{*}{$=$} &  	\multirow{2}{*}{$V_i^{I'} \left( \kappa_{ijk} (\sfrac32 V_j^\prime \cdot V_k^\prime + V_j^{I'} V_k^{I'}) + \sfrac32  c_{2i} \right) $} 
\\ &&&
\\ \hline 
\text{U}(1)^{I'} - \text{U}(1)^{J'} - \text{U}(1)^{J'} 		& 	\multirow{2}{*}{ $\sum_{p}\, \mathcal{N}\, p^{I'}\, p^{J'}\, p^{J'}$ } 	& \multirow{2}{*}{$=$} &  	\multirow{2}{*}{ $V_i^{I'} \left( \kappa_{ijk} (\sfrac12 V_j^\prime \cdot V_k^\prime + V_j^{J'} V_k^{J'}) + \sfrac12  c_{2i} \right)$ }
\\
I' \neq J' & & &
\\ \hline
\multirow{2}{*}{$\text{U}(1)^{I'} - \text{U}(1)^{J''} - \text{U}(1)^{J''}$} & \multirow{2}{*}{$\sum_{p}\, \mathcal{N}\, p^{I'}\, p^{J''}	p^{J''}$}  & \multirow{2}{*}{$=$} &  \multirow{2}{*}{$	0$} 
\\ &&&
\\ \hline
\text{U}(1)^{I'}-\text{U}(1)^{J'} -\text{U}(1)^{K'}		& 	\multirow{2}{*}{ $\sum_{p}\, \mathcal{N}\, p^{I'}\, p^{J'}\, p^{K'}$ }	& \multirow{2}{*}{$=$} &  	\multirow{2}{*}{ $\kappa_{ijk} V_i^{I'} V_j^{J'} V_k^{K'}$ }
\\
I' \neq J' \neq K' \neq I' & & &
\\ \hline
\text{U}(1)^{I'}-\text{U}(1)^{J''} -\text{U}(1)^{K''} 	& \multirow{2}{*}{ $\sum_{p}\, \mathcal{N}\, p^{I'}\, p^{J''}\,\, p^{K''}$}	& \multirow{2}{*}{$=$} & 	\multirow{2}{*}{ $0$ }
\\
J'' \neq K'' 	&	& 	&
\\ \hline
\end{array}
\renewcommand{\arraystretch}{\arrystrch}
\)
}
\end{center} 
\caption{\label{tb:E8xE8_GSmechanism}
This table presents the pure and mixed Abelian anomaly cancellation checks on the four-dimensional charged chiral spectrum obtained from line bundle Calabi-Yau compactifications of the supersymmetric E$_8\times$E$_8$ theory. 
All sums are only over those weights with $\cN=\cN(p;s) >0$. In addition, also the expressions with $' \rightarrow ''$, interchanging the observable and hidden sectors, are checked. 
}
\end{table}

\begin{table}[!hb]
\begin{center}
\scalebox{.95}{
\(
\renewcommand{\arraystretch}{1.1}
\begin{array}{|l||rcl|}
\hline 
\multicolumn{4}{|c|}{\text{\bf Supersymmetric \text{SO}(32) theory}}
\\ \hline\hline
\multirow{2}{*}{$\text{grav}-\text{grav}-\text{U(1)}^I$}  & \multirow{2}{*}{$-\sum_{p}\, \mathcal{N}\, p^{I}$}	 &\multirow{2}{*}{$=$}&	 \multirow{2}{*}{$V_i^I \left( 4 \kappa_{ijk} V_j^I V_k^I + \sfrac12 c_{2i} \right)$}
\\ &&&
\\ \hline 
\multirow{2}{*}{$G_x-G_x-\text{U(1)}^a$} & -\sum_\rep{R}\, n_x(\rep{R})\, \ell (\rep{R}_x)\, q^{a}_\rep{R}  &=& \sfrac13\, \sum_I t^I_a   \kappa_{ijk} V_i^I V_j^I V_k^I +
\\
& &&  + t^a \cdot V_i \Big( \sfrac16\, c_{2i} + \,  \sum_{w\in\rep{r}_x} \ell (\rep{r}_x)\, (w\cdot V_j)(w\cdot V_k) \Big)
\\ \hline 
\multirow{2}{*}{$\text{U}(1)^{I} - \text{U}(1)^{I} - \text{U}(1)^{I}$} & \multirow{2}{*}{$-\sum_{p}\, \mathcal{N}\, p^{I}p^{I}p^{I}$} 	&\multirow{2}{*}{$=$}&  	\multirow{2}{*}{$V_i^I \left( 4 \kappa_{ijk} V_j^I V_k^I + \sfrac12 c_{2i} \right)$} 
\\ &&&
\\ \hline 
\text{U}(1)^{I} - \text{U}(1)^{J} - \text{U}(1)^{J} 		& 	\multirow{2}{*}{ $-\sum_{p}\, \mathcal{N}\, p^{I}\, p^{J}\, p^{J}$ } 	& \multirow{2}{*}{=} & 	\multirow{2}{*}{ $V_i^{I} \left( \kappa_{ijk} \left( \sfrac13 V_j^I V_k^I + V_j^{J} V_k^{J} \right) + \sfrac16  c_{2i} \right)$ }
\\
I \neq J & & &
\\ \hline
\text{U}(1)^{I}-\text{U}(1)^{J} -\text{U}(1)^{K} 	& \multirow{2}{*}{ $\sum_{p}\, \mathcal{N}\, p^{I}\, p^{J}\,\, p^{K}$}	& \multirow{2}{*}{$=$} & 	\multirow{2}{*}{ $0$ }
\\
I \neq J \neq K \neq I 	&	& 	&
\\ \hline
\end{array}
\renewcommand{\arraystretch}{\arrystrch}
\)
}
\end{center} 
\caption{\label{tb:Spin32_GSmechanism}
This table presents the pure and mixed Abelian anomaly cancellation checks on the four-dimensional charged chiral spectrum obtained from line bundle Calabi-Yau compactifications of the supersymmetric SO(32) theory. 
All sums are only over those weights with $\cN=\cN(p;s) >0$.}
\end{table}

\newpage 

\begin{table}[t]
\begin{center} 
\scalebox{.95}{
\(
\renewcommand{\arraystretch}{1.1}
\begin{array}{|l||rcl|}
\hline 
\multicolumn{4}{|c|}{\text{\bf Non-supersymmetric SO(16)$\times$SO(16) theory}}
\\ \hline\hline
\multirow{2}{*}{$\text{grav}-\text{grav}-\text{U(1)}^{I'}$} & \multirow{2}{*}{$\sum_{p}\, \mathcal{N}\, p^{I'}$}	 &\multirow{2}{*}{$=$}&	 \multirow{2}{*}{$V_i^{I'} \left( \kappa_{ijk} \left( 3 (V_j^\prime \cdot V_k^\prime) - 4 V_i^{I'} V_j^{I'} V_k^{I'} \right) + 2 c_{2i} \right) $}
\\ &&&
\\ \hline 
\multirow{2}{*}{$G_{x'}-G_{x'}-\text{U(1)}^{a'}$} & \sum_\rep{R}\, n_{x'}(\rep{R})\, \ell (\rep{R}_{x'})\, q^{a'}_R   &=& 	
- \sfrac13 \sum_{I'}t_{a'}^{I'}\, \kappa_{ijk} V_i^{I'} V_j^{I'} V_k^{I'} + t^{a'} \cdot V^\prime_i \Big(  \sfrac13 c_{2i} + 
\\
& & 	&	+ \sfrac12 \kappa_{ijk}\, V_j^\prime \cdot V_k^\prime - \sum_{w\in \rep{r}_{x'}} \ell (\rep{r}_{x'})\, (p\cdot V^\prime_j)(p\cdot V^\prime_k) \Big)
\\ \hline
\multirow{2}{*}{$G_{x'}-G_{x'}-\text{U(1)}^{a''}$} & \sum_\rep{R}\, n_{x'}(\rep{R})\, \ell (\rep{R}_{x'})\, q^{a''}_R   &  =  & -\sfrac13\, \sum_{I''} t^{I''}_{a''}   \kappa_{ijk} V_i^{I''} V_j^{I''} V_k^{I''} - t^{a''} \cdot V_i^{\prime\prime} \Big( \sfrac16\, c_{2i} + 
\\
& &&  + \,  \sum_{w\in\rep{r}_{x'}} \ell (\rep{r}_{x'})\, (w\cdot V^\prime_j)(w \cdot V^\prime_k) \Big)
\\ \hline
\multirow{2}{*}{$\text{U}(1)^{I'} - \text{U}(1)^{I'} - \text{U}(1)^{I'}$} & \multirow{2}{*}{$\sum_{p}\, \mathcal{N}\, p^{I'}p^{I'}p^{I'}$} 	&\multirow{2}{*}{$=$}&  	\multirow{2}{*}{$V_i^{I'} \left( \kappa_{ijk} (\sfrac32 V_j^\prime \cdot V_k^\prime - 3 V_j^{I'} V_k^{I'}) + c_{2i} \right) $} 
\\ &&&
\\ \hline 
\text{U}(1)^{I'} - \text{U}(1)^{J'} - \text{U}(1)^{J'} 		& 	\multirow{2}{*}{ $\sum_{p}\, \mathcal{N}\, p^{I'}\, p^{J'}\, p^{J'}$ } 	&\multirow{2}{*}{=}&  	\multirow{2}{*}{ $V_i^{I'} \left( \kappa_{ijk} (\sfrac12 V_j^\prime \cdot V_k^\prime - \sfrac13 V_j^{I'}  V_k^{I'}) + \sfrac13  c_{2i} \right)$ }
\\
I' \neq J' & & & 
\\ \hline
\multirow{2}{*}{$\text{U}(1)^{I'} - \text{U}(1)^{J''} - \text{U}(1)^{J''}$} & \multirow{2}{*}{$\sum_{p}\, \mathcal{N}\, p^{I'}\, p^{J''}	p^{J''}$}   &\multirow{2}{*}{$=$} &  \multirow{2}{*}{$- V_i^{I'} \left( \kappa_{ijk} \left( \sfrac13 V_j^{I'} V_k^{I'} + V_j^{J''} V_k^{J''} \right) + \sfrac16  c_{2i} \right)$} 
\\ &&&
\\ \hline
\text{U}(1)^{I'}-\text{U}(1)^{J'} -\text{U}(1)^{K'}		& 	\multirow{2}{*}{ $\sum_{p}\, \mathcal{N}\, p^{I'}\, p^{J'}\, p^{K'}$ }	&\multirow{2}{*}{=} &   	\multirow{2}{*}{ $\kappa_{ijk} V_i^{I'} V_j^{J'} V_k^{K'}$ }
\\
I' \neq J' \neq K' \neq I' & &  & 
\\ \hline
\text{U}(1)^{I'}-\text{U}(1)^{J''} -\text{U}(1)^{K''} 	& \multirow{2}{*}{ $\sum_{p}\, \mathcal{N}\, p^{I'}\, p^{J''}\,\, p^{K''}$}	& \multirow{2}{*}{=} &  	\multirow{2}{*}{ $0$ }
\\
J'' \neq K'' 	&	& & 
\\ \hline
\end{array}
\renewcommand{\arraystretch}{\arrystrch}
\)
}
\end{center} 
\caption{\label{tb:nonSUSY_GSmechanism}
This table presents the pure and mixed Abelian anomaly cancellation checks on the four-dimensional charged chiral spectrum obtained from line bundle Calabi-Yau compactifications of the non-supersymmetric SO(16)$\times$SO(16) theory.  
All sums are only over those weights with $\cN=\cN(p;s) >0$. 
(In addition, also the expressions with $' \rightarrow ''$, interchanging the observable and hidden sectors, are checked.) 
}
\end{table}

\subsection{Analysis of the phenomenology}\label{sc:WorkingDefinitions}

Next, we introduce a couple of definitions to characterize smooth models and their 4D spectra. These will be used in the subsequent section to present and analyze our model-building results.

\subsubsection*{Model classification criteria}

We search for models that are close to the MSSM or the SM depending on whether model building is performed in the supersymmetric or non-supersymmetric context.
For this purpose, a smooth model in 4D (constructed on a CY $X$) is solely defined by
\items{
\item the full gauge group, i.e.\ its non-Abelian gauge symmetries and U(1)'s,
\item the chiral massless spectrum (bosons \& fermions),
\item its line bundle vectors.
}
At the next level, which goes beyond the scope of the present work, further elements of the effective field theory analysis in four dimensions, such as admissible D- and F-flat VEV configurations and (Yukawa) couplings, could be naturally added to the main definition.

In this paper, we have chosen to obtain (MS)SM physics through an SU(5) GUT theory. Therefore, we introduce the notion of GUT-like model for a smooth model when
\items{
\item there is an unbroken SU(5) GUT group allowed by the line bundle background 
\item there are representations available that admit the interpretation of SM quarks and leptons, i.e.\ $\rep{5}$- and $\crep{10}$-plets in the SU(5) case. 
}
Also the definition of GUT-like models can be naturally extended to SO(10) or E$_6$ GUTs, which we do not consider in the present work.
Analogously, we shall call a model (MS)SM-like when
\items{ 
\item the unbroken gauge group contains the SM gauge group $G_{SM}=$SU$(3)\times$SU$(2)\times$U$(1)_Y$
\item there is a net number of three chiral generations of quarks and leptons and at least three chiral singlets under $G_{SM}$ to accommodate the right-handed neutrinos
\item all exotic fermions are singlets w.r.t.\ the SM gauge group.
}
Since (MS)SM-like models in the downstairs picture are obtained from an upstairs GUT-like model via a freely acting Wilson line which breaks the GUT gauge group non-locally, the standard SU(5) normalization of the non-anomalous hypercharge $Y$ is automatically obtained in all the line bundle models.

Finally, we refer to a GUT-like model as chiral exact if the chiral number of $\rep{5}$- and $\crep{10}$-plets is exactly $3\, |\Gamma|$, before modding out the freely acting symmetry $\Gamma$. In other words, in this case there are no additional vector-like $\rep{5}$-$\crep{5}$ or $\rep{10}$-$\crep{10}$ pairs in the chiral spectrum detected by the multiplicity operator~\eqref{MultiOp4D}. 
Similarly, we call a spectrum (MS)SM-like chiral exact when the chiral number of generations of quarks and leptons is precisely three.

There are a couple of things to notice about these definitions. First of all, by definition a chiral exact GUT- or (MS)SM-like model cannot have any exotic fermionic states charged under SU(5) or $G_{SM}$, respectively. On the other hand, it could well (and most probably will) have chiral non-Abelian representations transforming under a hidden group $\widetilde{G}$, which will be singlets under SU(5) or $G_{SM}$\,. In addition, there will be various singlets charged under the various (hidden) U(1)'s. 
Secondly, starting from a specific GUT-like theory in the upstairs picture, there are generically various (MS)SM-like theories with different downstairs spectra, corresponding to the various ways in which the hidden gauge group can be broken and accordingly branched. In addition, there may be inequivalent ways to embed $G_{SM}$ inside SU(5), leading to distinct downstairs (MS)SM spectra as well.

Finally, in the present context, the multiplicities of 4D states are defined from the perspective of the GUT or $G_{SM}$ group. This means that fields that transform in a $d$-dimensional representation of an additional hidden gauge group are counted $d$ times. For example, consider a state $4\,(3,1;\,3)$ where 4 is the multiplicity of this smooth state as computed by the multiplicity operator, $(3,1)$ is the representation after branching the GUT group to $G_{SM}$ (suppressing U(1) charges) and $(3)$ is the dimension of the representation under some hidden surviving SU(3) symmetry. Using the above definitions, such a state would correspond to $N=4\cdot 3 = 12$ times the SM state $(3,1)$ from our four-dimensional perspective.

\subsubsection*{Inequivalent models}

After we have generated a large collection of GUT-like or (MS)SM-like models on a certain geometry we want to get a crude feeling for their phenomenological properties and analyze them statistically. We have to check whether all the models (not only GUT- or MSSM-like) we have generated on a given geometry are really distinct. Indeed, having exactly the same model appear over and over again in our list might lead to a misguided interpretation of statistical results. In particular, ignoring such equivalences in the spectra could have the effect that not all smooth models appear with equal probabilities in our statistics. For instance, a bundle background corresponding to a dense $K$-matrix is for sure much more unlikely to be constructed than another bundle with a sparse $K$-matrix. Moreover, since there are various non-linear and iterative steps in our scans, such inequivalence tests are crucial. 

In fact, even before that, we have to recall from Section~\ref{sc:ExamplesCYs} that there are geometries with different configuration matrices which are nevertheless identical on the level of the geometrical input data we make use of, i.e.\ intersection numbers and second Chern classes. For such groups of manifolds, like e.g.\ CICY~6715, 6788, 6836, 6927, we have only scanned over and investigated their model building potential once.\footnote{It is well possible that under more detailed phenomenological analyses models on such seemingly equivalent geometries might behave differently.}

There are two approaches to define in which cases two models are considered inequivalent on a given manifold: 
\begin{enumerate}[i.)]
 \item Two models are considered inequivalent when they have different chiral spectra up to their U(1) charges. \label{it:classificationApproach1}
 \item Two models are considered inequivalent when they are described by different line bundles.\label{it:classificationApproach2}
\end{enumerate}
The first approach has often been employed when scanning phenomenological orbifold models. In that context, it was avoided to explicitly compare U(1) charges, since in that case one has to consider all possible bases for these Abelian gauge symmetries, which is a computationally cumbersome procedure.

In the second approach we consider line bundle vectors to be distinct when there are no obvious symmetries that can map the two line bundle backgrounds onto each other. Hence, one should be careful to filter out as many symmetries as possible. There are basically two types of permutation symmetries to be considered in this context: Permutation of bundle vector entries $V_k^I\leftrightarrow V_k^J$ and permutation of fluxes $V_i \leftrightarrow V_j$ on divisors $D_i\,,\,D_j$\, which have the same Chern classes and whose intersection numbers with each other and all other divisors are also the same.
Despite taking such permutation symmetries into account, huge lists of inequivalent models are generated already in the upstairs picture. That is why in practice, before applying the freely acting symmetry, we have only kept lists of inequivalent models 
containing an SU(5) gauge group, 
see also the relevant statistics of section \ref{sc:SU5_statistics}.

Albeit seemingly unlikely, it happens quite frequently that two distinct sets of line bundle vectors lead to an identical (massless) spectrum. Because of this, the classification approach \ref{it:classificationApproach1}.) leads to much fewer inequivalent models. However, even if the low-energy spectra of two models with distinct line bundles are identical, their detailed phenomenology might still be different, since, for example, their (Yukawa) couplings are most likely distinct. For this reason we should consider two models to be inequivalent when either of the two approaches classify them as being inequivalent.

In the upstairs picture, approach \ref{it:classificationApproach2}.) implies approach \ref{it:classificationApproach1}.). So it is enough to keep a list of inequivalent line bundle backgrounds on a given geometry.
Concerning the equivalence between downstairs spectra, there is a subtlety to be noted. Since the Wilson line projection condition is defined modulo lattice vectors, it does not allow for a good classification of distinct models. On the other hand, we know from our previous discussion that the same upstairs smooth model generically corresponds to various distinct downstairs spectra. It is thus necessary to combine both approaches \ref{it:classificationApproach1}.) and \ref{it:classificationApproach2}.) to classify models in the downstairs picture according to their bundle realization as well as their downstairs massless spectrum.

\subsection{Generating line bundles on specific smooth manifolds}\label{sc:BundleGenerationExamples}

In this section we describe our model building and classification procedures for a couple of geometries (in particular those presented in Section~\ref{sc:ExamplesCYs}) in a bit more detail. We stress though, that our methods are more general and can in principle be applied to any smooth manifold. The purpose of this section is to illustrate these methods in concrete examples and prepare the setting for the actual model scans. On the other hand, we would also like to emphasize that it is sometimes beneficial to modify or by-pass certain steps. In particular, the structure of the Schoen manifold leads to various simplifications in the model building process. 

\subsubsection*{Generating line bundles and classifying models on favorable CICYs}

In this paragraph we focus on a specific type of favorable CICY geometries, i.e.\ CICYs 7862, 7447, 7487 and 5302, which contain intersecting K3 manifolds. Using the intersection numbers of Table~\ref{tb:CICYs_summary}, their Bianchi identities written in terms of the Gram matrix $K$ can be compactly summarized as
\equ{ 
\vert \ge_{ijk} \vert K_{jk} = -24~, 
}
for $1\leq i \leq h_{11}$ using the standard definition of the totally antisymmetric tensor $\ge_{ijk}$\,. 
Consequently, these equations do not restrict the diagonal entries $K_{ii}=V_i^2$ of the Gram matrix. Hence, only the tree-level DUY equations impose constraints on the norms of the line bundle vectors.

\begin{table}[t]
\[
\renewcommand{\arraystretch}{1.3}
\begin{array}{|c|c|}
\hline 
\multicolumn{2}{|c|}{\text{\bf Symmetric K3-like \textbf{CICY~7862} }}
\\ 
\text{\bf Bianchi identities} & \text{\bf Tree-level DUY}
\\ \hline\hline 
K_{14} = -12-K_{12}-K_{13}   &	v_1\, K_{11} = 12 v_4 + (v_4-v_2)K_{12} + (v_4-v_3)K_{13}
\\ \hline
K_{23} = K_{14}              & 	v_2\, K_{22} = 12 v_3 + (v_3-v_1)K_{12} + (v_3-v_4)K_{13}
\\ \hline
K_{24} = K_{13}   			&   v_3\, K_{33} = 12 v_2 + (v_2-v_4)K_{12} + (v_2-v_1)K_{13}
\\ \hline
K_{34} = K_{12}    			& 	v_4\, K_{44} = 12 v_1 + (v_1-v_3)K_{12} + (v_1-v_2)K_{13}
\\ \hline 
\end{array}
\renewcommand{\arraystretch}{\arrystrch}
\]
\caption{\label{tb:CICY7862_BIsDUY}
This table gives an explicit solution of the BIs in terms of $K_{ij}$ for CICY~7862  (with $h_{11}=4$) from the special class of favorable symmetric K3-like CICYs. In the second column the norms of bundle vectors are constrained via the DUY equations. There are two free entries $K_{ij}$ ($K_{12}$ and $K_{13}$) that remain unconstrained by both the BIs and the DUY equations. We have abbreviated $\text{Vol}(D_i)=v_i$\,.}
\end{table}

Let us demonstrate this observation explicitly for such a K3-like symmetric CY with $h_{11}=4$, CICY~7862, also called the tetra-quadric (cf.\ Table~\ref{tb:CICY7862_BIsDUY}): As outlined in Section~\ref{sc:KahlerConeGeneration} we first generate a large collection of points inside the K\"ahler cone. Next, we use the linear system of equations imposed by the Bianchi identities to solve $K_{14}, K_{24}, K_{24}, K_{34}$ in terms of $K_{12}$ and $K_{13}$, cf.\ the first column of Table~\ref{tb:CICY7862_BIsDUY}. In the right column, we then incorporate the relations derived from the Bianchi identities into the DUY conditions~\eqref{DUY_sol}. This results in only two independent Gram matrix entries, $K_{12}, K_{13}$, out of the initially $\sfrac{4 \cdot 5}{2}=10$ independent components; all other entries are given by the two off-diagonal $K_{ij}$ and ratios of divisor volumes Vol$(D_i)$\,. 

Subsequently, we need to go through the various points inside the K\"ahler cone and search for particular values of $K_{12}, K_{13}$, which support non-negative even integer $K_{ii}$ for all $i=1,\ldots,h_{11}$. 
This last step is far less trivial than one might initially think: The volumes Vol$(D_i)$ are not some randomly adjustable parameters, but are (at least) related through the inequalities stated in Table~\ref{tb:CICYs_KahlerCone}. For instance, on the tetra-quadric no volume can become very big compared to the other three, since it is maximally constrained by the sum of the other volumes.

In total, we see that using the strategy presented in section~\ref{sc:GramMatrixGeneration} with some elementary linear algebra, we have managed to reduce a highly coupled non-linear Diophantine system of equations in the bundle vectors entries $V_i^I$ to simply scanning over integer values for the two unconstrained Gram matrix entries and performing some simple combinatorics. Since all the necessary steps have been performed in a theory-independent way, one can use the procedure (described in section~\ref{sc:BundleVectorsGeneration}) to generate vectors on the appropriate lattice for any of the heterotic theories.

\begin{table}[t]
\begin{center}
\renewcommand{\arraystretch}{1.5}
\begin{tabular}{|c||c|}
\hline
{\bf CICY \#}	&	{\bf Necessary conditions to be inside the K\"ahler cone}
\\ \hline\hline
{\bf 7862}				&	Vol$(D_i) < \text{Vol}(D_j) + \text{Vol}(D_k) + \text{Vol}(D_l$) for $i,j,k,l \in \lbrace 1,\ldots,4\rbrace$ all different
\\ \hline
{\bf 7491,\,7522}		&	Vol$(D_a)+\text{Vol}(D_b) < \text{Vol}(D_4) < 2 \sum_{c=1}^3 \text{Vol}(D_c)$ for $a,b \in \lbrace 1,2,3 \rbrace$
\\ \hline
{\bf 7447,\,7487}		&	Vol$(D_i) < \text{Vol}(D_j) + \text{Vol}(D_k) + \text{Vol}(D_l$) for $i,j,k,l \in \lbrace 1,\ldots,5\rbrace$ all different
\\ \hline
{\bf 6770}  		    &	$\text{Vol}(D_a) <  \text{Vol}(D_b) + \text{Vol}(D_4) + \text{Vol}(D_5)$ for $a \neq b \in \lbrace 1,2,3 \rbrace$
\\
      					&	$2 \text{Vol}(D_a) < \sum_{b=1}^3 \text{Vol}(D_b)$ for $a=4,5$
\\ \hline	
{\bf 6715,\,6788 }	    &	$\text{Vol}(D_a) +  \text{Vol}(D_b) <  \text{Vol}(D_5) <  \sum_{c=1}^4 \text{Vol}(D_c)$ 
\\
{\bf 6836,\,6927}       &	$\text{Vol}(D_a) < \sum_{a \neq c=1}^4  \text{Vol}(D_c)$ for $a,b \in \lbrace1,\ldots,4\rbrace$  	
	  
\\ \hline
{\bf 6732,\,6802} 	   	&	$\text{Vol}(D_1) + \text{Vol}(D_2) < \text{Vol}(D_5) < \sum_{c=1}^4 \text{Vol}(D_c)$
\\
{\bf 6834,\,6896 }    	&	$\text{Vol}(D_a) < \sum_{a \neq c=1}^4 \text{Vol}(D_c)$  for $a \in \lbrace 1,\ldots,4 \rbrace$
\\
					  	&	$\text{Vol}(D_b) <  \text{Vol}(D_5)$  for $b=3,4$
\\ \hline
{\bf 6225}				&	$\text{Vol}(D_a) + \text{Vol}(D_b) < \text{Vol}(D_4) + \text{Vol}(D_5)$ for $a,b \in \lbrace 1,2,3 \rbrace$
\\
						&	$\text{Vol}(D_4) + \text{Vol}(D_5) < 3 \sum_{c=1}^3 \text{Vol}(D_c)$
\\
     					&	$\text{Vol}(D_d) < \sum_{d \neq c=1}^5 \text{Vol}(D_c)$ for $d = 4,5$
\\ \hline 
{\bf 5302}   		    &	$\text{Vol}(D_i) < \text{Vol}(D_j) + \text{Vol}(D_k) + \text{Vol}(D_l$) for $i,j,k,l \in \lbrace 1,\ldots,6\rbrace$ all different
\\ \hline        
\end{tabular}
\renewcommand{\arraystretch}{1}
\end{center}
\caption{\label{tb:CICYs_KahlerCone}
This table presents necessary conditions on the divisor volumes Vol$(D_i)$ in order to be inside the K\"ahler cone for the various geometries of Table~\ref{tb:CICYs_summary}. These conditions are derived using the definition of divisor volumes \eqref{VolumesWithDilaton} and the appropriate intersection numbers. For all favorable CICYs presented here the volumes of curves dual to the divisors $D_i$ and of the full Calabi-Yau are positive iff $a_i > 0$\,.
}
\end{table}

To classify inequivalent models in the upstairs picture we employ classification approach~\ref{it:classificationApproach1}.) as discussed in Section~\ref{sc:WorkingDefinitions}. It is crucial to filter out various permutation symmetries in the bundle data used to classify model equivalence:
To systematically filter out permutation symmetries among vector entries $V_k^I \leftrightarrow V_k^J$ we have implemented a geometry-independent routine.
In contrast, one should be more careful when considering the second type of permutation symmetry, i.e.\ interchanging fluxes $V_i \leftrightarrow V_j$ on divisors whose intersection numbers and Chern classes are the same. Since the statement is obviously geometry-depended, one needs to perform a pre-analysis for each geometry in order to see for which divisors this is the case. For instance, we immediately see that in our specific example, CICY~7862, all divisors are K3-surfaces, which yields $4!=24$ different possibilities to rearrange the exact same bundle background by permuting all four $V_i$\,. For a faithful phenomenological analysis only one of those equivalent backgrounds has to be stored.

Fortunately, for the limited set of favorable CICYs we consider in this paper, there is a clear pattern for the various divisors Vol$(D_i)$ and their triple intersection numbers $\gk_{ijk}$\,. Concretely, we find that for most of these geometries, many (if not all) divisors do not have any self-intersections (i.e. $D_i^2 =0$) at all. For this subset of divisors the geometry looks essentially identical to the K3-like CICYs we have just investigated. For the self-intersecting divisors (i.e. $D_i^2 \neq 0$) the situation deviates from our previous discussion. Usually there are one or two self-intersecting divisors (which tend to have the same Chern classes and intersection numbers among themselves).
This motivates the following strategy: We group the CICYs we want to scan according to the number of self-intersecting divisors they have, as done in Table~\ref{tb:CICYs_classification}. For a given group of manifolds the properties under the exchange of fluxes on divisors with the same $c_{2i}$ and $\gk_{ijk}$ are mostly identical or at least very similar.

\begin{table}
\begin{center}
\renewcommand{\arraystretch}{1.2}
\begin{tabular}{|c|c|c||c|}
\hline
\multicolumn{3}{|c||}{\bf Favorable \textbf{CICY}s}		&	{\bf Number of self-}
\\
$h_{11}=4$	& 	$5$ 		& 	$6$			&   {\bf intersecting divisors}
\\ \hline\hline 
\multirow{2}{*}{{7862}} 	& 	{7447}, {7487}						&	\multirow{2}{*}{{5302}}	&	\multirow{2}{*}{0}
\\
								&	{6770}									&									&
\\ \hline
\multirow{2}{*}{{7491}, {7522}}		&	{6715}, {6788}, {6836}, {6927}		&	&	\multirow{2}{*}{1}
\\
								&	{6732}, {6802}, {6834}, {6896}						&	&
\\ \hline
								&	{6225}											&	\phantom{{7491}, {7522}} &	2
\\ \hline							
\end{tabular}
\renewcommand{\arraystretch}{1}
\end{center}
\caption{\label{tb:CICYs_classification}
The favorable CICYs considered in our scans, are grouped according to the number of self-intersecting divisors they possess. This classification is helpful to filter out permutation symmetries among equivalent bundle vectors supported on these divisors.
}
\end{table}

\subsubsection*{Generating line bundles on the Schoen manifold}

The Schoen manifold was introduced in Section~\ref{sc:Schoen}. In order to generate chirality, we switch on magnetic fluxes, encoded by line bundle vectors, $B_1, B_2, B_3$, on the inherited divisors $R_1, R_2, R_3$\,, respectively. On the exceptional cycles $E_r$ and $\widetilde{E}_r$ we use fluxes $V_r$, $\widetilde{V}_r$\, respectively. The multi-index $r$ was defined in Section~\ref{sc:Schoen}. To satisfy the flux quantization, we require that \cite{Nibbelink:2012de} 
\equ{ 
B_1\,,\, B_2\,,\, B_3 \in \Lambda~, 	\qquad	 2\,V_r\,,\, 2\, \widetilde{V}_r \in  \Lambda\,,
}
with $\Lambda$ given in \eqref{GaugVectors_on_lattice}. Using the intersection numbers from Table~\ref{tb:Schoen_info} the anomaly cancellation requirement (without the presence of non-perturbative brane effects) in four dimensions amounts to 
\begin{subequations}
\equ{ 
\sum_r (V_r)^2 = 12 + 2 B_2 B_3~,
\qquad  
\sum_{r} (\widetilde{V}_{r})^2 = 12 + 2 B_1 B_3~,
\\[1ex] 
B_1 \cdot V_r = 0~, 
\qquad  
B_2 \cdot \widetilde{V}_{r} = 0~, 
\qquad 
B_1 \cdot B_2 = 0~.
}
\end{subequations}
The DUY conditions are more involved compared to the CICY case, as can be seen from Table~\ref{tb:Schoen_info}. Roughly, they can be characterized as $\text{Vol}(R_i) > 4 \text{Vol}(E_r) > 0$ (and similarly for the divisors $\widetilde{E}_r$). 

To simplify the analysis we aim to satisfy the tree-level DUY equations for the inherited divisors independent of the exceptional ones. Thus the magnetic fluxes are taken to satisfy
\equ {
B_1 + B_2 + B_3 = 0\,.
}
As a convenient ansatz we further assume that sets of two bundle vectors $V_{r}$ (analogously for $\widetilde{V}_{r}$) are always equal or opposite, in such a way that the equivariant identifications are respected and the DUY equations~\eqref{DUY1} are satisfied simultaneously:
\equ {
V_{r_1 r_2 r_3} = (-1)^{r_2+r_3}\, V_{r_100}\,.
}
In particular, by choosing to satisfy the tree-level DUY separately for the inherited and exceptional divisors, it is guaranteed that K\"ahler parameters $a_i,b_r,\tilde b_r$ exist such that all relevant volumes are positive.

Note that in addition, the whole construction (i.e.\ consistency conditions and spectrum) is invariant under exchange of fluxes on divisors with permutation symmetries as explained above, e.g.\ $B_1 \leftrightarrow B_2$ together with $V_r \leftrightarrow \widetilde{V}_r$\,. As with the CICYs discussed in the previous paragraph, one also needs to take such permutation symmetries into account in order to avoid an enormous over-counting.

Compared to the favorable CICYs reviewed in the previous section, we find that the Schoen manifold has some convenient features in the non-integral basis defined in Table~\ref{tb:Schoen_info}: in contrast to the favorable CICYs, where all scalar products among different vectors appear in Bianchi identities, most of the off-diagonal entries of the Gram matrix for this manifold remain fully unconstrained. Especially computationally, this has important consequences in random bundle generation, since we only need to generate a handful of $V_r$'s of a given norm, which is significantly faster than any known algorithm for generating vectors of fixed inner products. Furthermore, the simple ansatz for the DUY equations resolves the seemingly complicated Mori/K\"ahler cone conditions into very simple inequalities for the divisor volumes. For the favorable CICYs, in spite of having parameterized solutions to the Bianchi and the DUY conditions, we were not able to find in this set-up an analytic expression to be inside the K\"ahler cone; instead we have to go through the combinatorial exercise of the previous section. 

\subsection{Computer implementation}

The computer code that automates the various procedures described above will be made public in the future under the name 
\texttt{Compactifier}. This code can be thought of as a major extension to the \texttt{Orbifolder} package~\cite{Nilles:2011aj}, which enables the latter to also address smooth line bundle compactifications. It uses a lot of the group theoretical routines of the original \texttt{Orbifolder} code to perform model analyses, but all the routines to generate consistent line bundle model input data and to construct the chiral spectra have been written from scratch. Moreover, the anomaly checks of the \texttt{Orbifolder} have been extended to include the non-universal Green-Schwarz mechanism on smooth Calabi-Yaus with a collection of axions for all three heterotic theories.  

Since the chiral spectra are computed using index theorems, a smooth model construction is in practice independent of $h_{11}$ and can be implemented very efficiently.
On the other hand, the group theoretical routines of the original \texttt{Orbifolder} have been optimized for fast model building analysis independent of the particular string setting under consideration. This means that the construction of a smooth model together with the basic consistency checks performed takes in average 0.005 seconds.
In our specific setting, where we choose to obtain the (MS)SM spectrum via an intermediate SU(5) GUT-like theory, it is actually of great computational benefit to first determine the unbroken gauge group and proceed with the computation of the chiral spectrum only if the smooth model under construction contains an SU(5) group factor.

In order to ensure that this code is not specific to one particular type of smooth Calabi-Yau space, we have detached the definition of the geometry from the actual code. The information is stored in a so-called geometry file that contains basic topological data of a Calabi-Yau geometry as described in Section~\ref{sc:ExamplesCYs}, like its intersection numbers $\gk_{ijk}$ and the values of the the integrated second Chern classes $c_{2i}$. Line bundle data can either be entered by hand on the level of a bundle file that specifies the line bundle vectors, or can be randomly generated using the procedures described in this section. 

The results of such model searches using this code are presented in the next section for all three heterotic string theories.

\section{Model searches}
\label{sc:Models}

We now present our results for heterotic model-building on a selection of favorable CICYs and the Schoen manifold. For all scans on CICYs we discuss in this section, the Standard Model is obtained from an intermediate SU(5) GUT by modding out an order two freely acting symmetry of the underlying geometry. Scan routines were executed simultaneously for four core-weeks to construct more than $10^9$ smooth models from each theory on every different geometry.

Favorable CICYs with $h_{11}=1$ have not been considered here, since the tree-level DUY equations cannot be satisfied inside the K\"ahler cone. For CICYs with small $h_{11} = 2,3$, it turns out that the conditions \eqref{K_matrix_conditions} together with the Bianchi identities and the tree-level DUY equations are often over-constraining: 
At least inside the parameter range $|K_{ij}| \leq 40$ we have scanned over, we found almost no solutions to the Bianchi identities, for most of the favorable CICYs with Hodge number $h_{11} < 4$. For this reason we do not consider such very small $h_{11}$ CICY geometries in the following.

\subsection{Statistics of model searches}

Tables~\ref{tb:ScanOverviewE8xE8}, \ref{tb:ScanOverviewSO32} and \ref{tb:ScanOverviewSO16xSO16} present our results from model scans over the same set of favorable CICYs with $4 \leq h_{11}\leq 6$ and the Schoen manifold of the two supersymmetric E$_8\times$E$_8$ and SO(32) heterotic theories and the non-supersymmetric SO(16)$\times$SO(16) theory, respectively. The left column always lists the geometry we have considered. The number refers to the CICY label in the classification list~\cite{Candelas:1987kf,Braun:2010vc}.
Following the definitions of Section~\ref{sc:WorkingDefinitions}, the two sets of columns on the right give the total number of inequivalent models we have constructed on each geometry. Here we distinguish between upstairs and downstairs models, i.e.\ between SU(5) GUT-like models and the resulting (MS)SM-like models obtained after modding out the compatible freely acting symmetry. In particular, the fourth and sixth column gives the number of GUT-like and (MS)SM-like models respectively, which are chiral exact. 

As expected from our remarks in Section~\ref{sc:WorkingDefinitions} the number of downstairs models is indeed (much) larger than the number of upstairs models. One the one hand, there are various possible group theoretical breakings for the hidden sector gauge group. This fact results in multiple models in the downstairs spectrum, which correspond to the same SU(5) GUT. In addition, it is often possible to embed the SM gauge group in more than one way inside the observable sector gauge group, by appropriate choices of a hypercharge generator $Y$, such that the desired GUT hypercharge normalization is preserved.

\subsubsection*{Geometry-based analysis} 

The results of these tables can be interpreted both from the perspective of different geometries or of the different theories. We begin with discussing some geometric features: 

The symmetric K3-like CICYs are most fruitful among the set of favorable CICYs we have considered in these scans. As outlined in section~\ref{sc:BundleGenerationExamples} this is due to the absence of a norm constraint on $V_i^2$ (i.e.\ on $K_{ii}$ for all $i$) coming from the BIs. Since this statement is solely based on geometric data, it is independent of the particular theory under consideration.

One might have expected that among similar geometries (e.g.\ symmetric K3-like CICY spaces, i.e.\ the first row of Table~\ref{tb:CICYs_classification} without self-intersections), the number of models would grow with increasing $h_{11}$, as the number of bundle vectors increases. Seemingly from the results under inspection this expectation is not verified. A possible reason is that we have scanned over the same duration for all geometries independently of $h_{11}$, even though, as remarked in Section~\ref{sc:RandomModels}, generating line bundle vectors for a given $K$-matrix scales polynomially in time with larger $h_{11}$. A rough estimate for instance, shows that one would need to scan three times longer on the CICY~5302 to produce as many models as on CICY~7862. 

Finally, we see that the Schoen manifold is a geometry which is particularly fruitful for the purpose of model building. As explained above, the main reason for this is that the Schoen manifold essentially only restricts norms of bundle vectors but leaves a lot of freedom otherwise.

\subsubsection*{Theory-based analysis} 

As far as we know the analysis presented here is novel in the sense that all three heterotic string theories are compared side-by-side for the first time when compactified on the same smooth geometries. We find that in our scans the E$_8\times$E$_8$ theory is the most fruitful for nearly every geometry (with the exception of the Schoen manifold, where SO(32) produces more models) we have scanned. The number of interesting GUT-like and SM-like models obtained within the SO$(16)\times$SO$(16)$ theory is comparable to the number of GUT- and MSSM-like models in the supersymmetric context.

Given the revived interest in the non-supersymmetric SO(16)$\times$SO(16) heterotic theory, let us concentrate on it in a bit more detail to understand the outcome more closely. In a previous version of this paper, we had restricted the line bundle vectors to lie on the lattice {\bf R}$_8\oplus${\bf R}$_8$, overlooking the fact that bundle vectors in {\bf S}$_8\oplus${\bf S}$_8$ are also compatible with flux quantization. This restriction had vetoed for the non-supersymmetric case many bundle vector solutions that are admissible in the supersymmetric theories. Consequently, there were thousand times less inequivalent models than in the E$_8\times$E$_8$ case on each geometry. Lifting this obsolete restriction we find that the number of semi-realistic SM-like models is very similar to their supersymmetric counterparts. A similar statistical outcome holds for the number of models with exotics. In particular, as can be seen from Table~\ref{tb:ScanOverviewSO16xSO16}, we are able to find some SU(5) GUT-like (and consequently SM-like) models which are chirally exact for either their fermionic or bosonic spectra and sometimes even for both.

\newpage

\begin{table}[!ht]
\begin{center}
\renewcommand{\arraystretch}{1.2}
\begin{tabular}{|c|c||c|c|c|c|}
\hline
\multicolumn{6}{|c|}{\bf Inequivalent SU(5) models for the $\boldsymbol{\text{E}_8\times\text{E}_8}$ theory on smooth CYs}
\\ \hline
\multicolumn{2}{|c||}{\bf Geometry}		 & 	  \multicolumn{2}{c|}{\bf Upstairs picture}   	&	\multicolumn{2}{c|}{\bf Downstairs picture}
\\
$h_{11}$				& 	(Name / CICY \#)		& 		GUT-like		& 	Chiral exact		&		MSSM-like	 & 	Chiral exact
\\ \hline\hline
$4$ &  Tetra-quadric (7862)   &  	245,387		& 	39,375	& 		1,571,972			&	328,445
\\ \hline
$4$ &  7491, 7522   &  2,099	 	& 	56		& 	17,928	&	764
\\ \hline
$5$ &  7447, 7487   & 177,359  & 	33,046		& 	1,106,276		&	352,458
\\ \hline
$5$ &  6770   &  110,823	 	& 	8,286		& 	816,098		&	86,975
\\ \hline
$5$ &  6715, 6788, 6836, 6927   & 	3,011  & 	239		& 	17,704		&	1,218
\\ \hline
$5$ &  6732, 6802, 6834, 6896   & 	27,898	  & 		823		& 	203,210		&	11,443
\\ \hline
$5$ &  6225   &  	2,016	 & 		0		& 		26,674		&	0
\\ \hline
$6$ &  5302   & 	154,352		& 	13,905	&	807,234		&	100,941
\\ \hline
$19$ &   Schoen  &  	355,745			 &		186,630		& 	2,631,198		&	1,682,159
\\ \hline
\end{tabular}
\renewcommand{\arraystretch}{1}
\end{center}
\caption{\label{tb:ScanOverviewE8xE8} 
This table provides results of our searches for MSSM-like models within the $\text{E}_8\times\text{E}_8$  heterotic theory on selected smooth Calabi-Yau manifolds.
The inequivalent SU(5) models with 6 generation GUTs become MSSM-like after modding out a freely acting $\mathbb{Z}_2$ Wilson line.
}
\end{table}

\begin{table}[!hb]
\begin{center}
\renewcommand{\arraystretch}{1.2}
\begin{tabular}{|c|c||c|c|c|c|}
\hline
\multicolumn{6}{|c|}{\bf Inequivalent SU(5) models for the SO(32) theory on smooth CYs}
\\ \hline
\multicolumn{2}{|c||}{\bf Geometry}		 & 	  \multicolumn{2}{c|}{\bf Upstairs picture}   	&	\multicolumn{2}{c|}{\bf Downstairs picture}
\\
$h_{11}$				& (Name / CICY \#)		& 		GUT-like		&	Chiral exact		&	MSSM-like	 & 	Chiral exact
\\ \hline\hline
$4$ &  Tetra-quadric (7862)   & 	159,510		&	841		& 	1,128,286		&	5,760
\\ \hline
$4$ &  7491, 7522   &  351 	& 	1	& 	2,956		&	8
\\ \hline
$5$ &  7447, 7487   &  69,669	 & 	3,385	& 	561,911		&	34,076
\\ \hline
$5$ &  6770   &  53,712	 & 	74	& 	410,830		&	547
\\ \hline
$5$ &  6715, 6788, 6836, 6927   &  407	& 	0	&  2,204		&	0
\\ \hline
$5$ &  6732, 6802, 6834, 6896   & 	6,017 	& 	2,655		& 	47,251	&	20,227
\\ \hline
$5$ &  6225   & 	 	0		& 	0	& 	0	&	0
\\ \hline
$6$ &  5302   &   2,598	& 	42	& 	15,326	&	232
\\ \hline
$19$ &   Schoen  	& 	493,114 		& 	207,644		& 	4,029,615	&	1,939,579
\\ \hline
\end{tabular}
\renewcommand{\arraystretch}{1}
\end{center}
\caption{\label{tb:ScanOverviewSO32} 
This table provides results of our searches for MSSM-like models within the SO(32) heterotic theory on selected smooth Calabi-Yau manifolds.
The inequivalent SU(5) models with 6 generation GUTs become MSSM-like after modding out a freely acting $\mathbb{Z}_2$ Wilson line.
}
\end{table} 

\newpage 

\begin{table}[!ht]
\begin{center}
\renewcommand{\arraystretch}{1.2}
%
%

%
%
%
\scalebox{.94}{
\begin{tabular}{|c|c||c|c|c|c|c|c|c|c|}
\hline
\multicolumn{10}{|c|}{\bf Inequivalent SU(5) models for the SO(16)$\boldsymbol{\times}$SO(16) theory on smooth CYs}
\\ \hline
\multicolumn{2}{|c||}{\bf Geometry}		 & 	  \multicolumn{4}{c|}{\bf Upstairs picture}   	&	\multicolumn{4}{c|}{\bf Downstairs picture}
\\
\multirow{2}{*}{$h_{11}$}				& \multirow{2}{*}{(Name / CICY \#)}		& 		\multirow{2}{*}{GUT-like}		&	\multicolumn{3}{c|}{Chiral exact}		&		\multirow{2}{*}{SM-like} 		& 	\multicolumn{3}{c|}{Chiral exact}	
\\
					&					&					&   Fermi & Scalar & Both 	& 					&  Fermi & Scalar & Both
\\ \hline\hline
$4$ &  Tetra-quartic (7862)   &  	209,743 	& 	281	& 	263		& 1 	&  1,575,098	& 	2,370	& 	2,000 	& 15	
\\ \hline
$4$ &  7491, 7522   &   1,873	& 	0	& 	1	&   0 	& 	14,651	& 	0	&	11	&	0
\\ \hline
$5$ &  7447, 7487   & 	28,209	 & 	901	& 	46	&  5	& 	149,143	& 	5,154 	&  377	&	52			
\\ \hline
$5$ &  6770   &   65,888	& 	173	& 	85	&  0	& 	437,327	& 	914	&	707	& 0
\\ \hline
$5$ &  6715, 6788, 6836, 6927   &  	120	 & 	7	&  0		&  0	& 	518	& 	89	& 	0	&	0			
\\ \hline
$5$ &  6732, 6802, 6834, 6896   &  460  	& 	33	&  0  	&  0	& 	3,119	& 	275	& 0	&	0			
\\ \hline
$5$ &  6225   &	  72	 &	0	 & 	0 	&  0	& 	483	& 	0	& 	0		&	0	
\\ \hline
$6$ &  5302   &		355 	 &	22	  & 0 	&  0	& 	1093 	& 	66	& 0	 & 0
\\ \hline
$19$ &   Schoen  	& 	456,594 	 &		5,169	  	& 	2,745 	&   	30	&  	3,002,353	  & 37,276	 &  21,955	&  237
\\ \hline
\end{tabular}}

\renewcommand{\arraystretch}{1}
\end{center}
\caption{\label{tb:ScanOverviewSO16xSO16} 
This table provides results of our searches for SM-like models within the $\text{SO}(16)\times\text{SO}(16)$ heterotic theory on selected smooth Calabi-Yau manifolds. The inequivalent SU(5) models with 6 generation GUTs become SM-like after modding out a freely acting $\mathbb{Z}_2$ Wilson line.
In upstairs and downstairs picture it is indicated how many models have no fermionic or scalar chiral exotics or whether it is fully exophobic at the chiral level (i.e.\ net number three of chiral SM families and no Higgs triplet, doublets).
}
\end{table}

\subsubsection*{Counting the number of states: Superfields versus bosons and fermions}

Concerning the number of chiral states, let us stress that the notion of exotics is very different depending on whether it is used in the supersymmetric or non-supersymmetric context. In order to describe the four-dimensional spectrum of one of the supersymmetric strings, we follow the convention to use superfields which include both bosonic and fermionic states. For compactifications of the SO(16)$\times$SO(16) theory, we adopt the convention of \cite{Blaszczyk:2014qoa} to display bosons and fermions separately since they are generically in very different representations. Hence, when comparing a supersymmetric spectrum to a non-supersymmetric one, one has to realize that even though the number of exotic superfields is smaller than the number of exotic fields in the non-supersymmetric theory, this might correspond to more exotics in the supersymmetric theories when bosons and fermions are considered separately. In particular, the scalar partners of the MSSM matter multiplets are considered as scalar exotics in the non-supersymmmetric context. Indeed, given the fact that we start with a smaller representation for the scalar fields in the ten-dimensional SO(16)$\times$SO(16) theory as compared to the supersymmetric theories, it is to be expected that there are less scalar exotics at the GUT- or SM-level. 

\subsubsection*{Total number of models and infinite sets}

We close the analysis of the chiral data with a remark on the total number of models we have found. First of all, we stress that these scans were not aimed to be exhaustive by any means. The duration of four weeks was mostly dictated by our requirement to obtain representative statistical results and not by some upper bound on the total number of inequivalent GUT-like/MSSM-like models we were able to spot. Furthermore, as explained in~\cite{Nibbelink:2015ixa} for smooth compactifications with line bundles there seems to be no clear theoretical bound on the total number of models supported on a given manifold inside the validity of the supergravity approximation. In that sense, we think that the very notion of exhaustive scans is not well-defined on those smooth manifolds and we have thus concentrated primarily on generic statistical findings.

\subsection{Analysis beyond the chiral spectrum}

\begin{table}[t]
\renewcommand{\arraystretch}{1.1}
\begin{center}

 \begin{tabular}{|c||c|c|c|c}
 \hline
{\bf Geometry}			& {\bf Heterotic} 	& {\bf  Higgs candidate}		&  {\bf Vector-like exotics }			\\
(Name / CICY \#)				& {\bf theory}		& { \#($\rep{5}$-$\brep{5}$) ${\geq1}$} 	& {(extra $\rep{5}$-$\brep{5}$, $\rep{10}$-$\brep{10}$ pairs)}	\\
  \hline\hline
   \multirow{2}{*}{Tetra-quadric (7862)}				& E$_8\times$E$_8$		& 100\%				& 0\%							\\
						& SO(32)			& 95.7\%			& 0\%							\\
 \hline
  \multirow{2}{*}{7491, 7522}			& E$_8\times$E$_8$		& 100\%				& 0\%							\\
						& SO(32)			& 100\%				& 0\%							\\
  \hline
  \multirow{3}{*}{7447, 7487}			& E$_8\times$E$_8$		& 92.4\%			& 7.0\%							\\
						& SO(32)			& 98.2\%			& 1.7\%							\\
  \hline
  \multirow{2}{*}{6770}				& E$_8\times$E$_8$		& 41.9\%			& 58.0\%						\\
						& SO(32)			& 100\%				& 0\%							\\
  \hline
  \multirow{1}{*}{6715, 6788, 6836, 6927}	& E$_8\times$E$_8$		& 100\%				& 0\%							\\
  \hline
  \multirow{2}{*}{6732, 6802, 6834, 6896}	& E$_8\times$E$_8$		& 58.4\%			& 40.9\%						\\
						& SO(32)			& 83.2\%			& 16.6\%						\\
  \hline
  \multirow{2}{*}{5302} 			& E$_8\times$E$_8$		& 92.1\%			& 3.3\%							\\
						& SO(32)			& 69.0\%			& 0\%							\\
  \hline
  \end{tabular}
\end{center}
\renewcommand{\arraystretch}{1}
\caption{\label{tab:FullSpectrumAnalysis} 
In this table we indicate the percentage of chiral exact SU(5) GUT-like models obtained from the supersymmetric heterotic string theories with additional non-chiral states which are charged under the SU(5). The Higgs multiplets can arise from vector-like $\rep{5}$-$\crep{5}$ pairs. The other non-chiral states constitute vector-like exotics, which can also come from additional $\rep{10}$-$\crep{10}$ pairs.
}
\end{table}

In this subsection we investigate smooth Calabi-Yau compactifications with line bundles beyond the chiral spectrum. This analysis is important since vector-like $\rep{5}$-plets can provide potential Higgs candidates while solving the doublet-triplet splitting problem, see e.g.\  \cite{Blaszczyk:2015zta} for a discussion of this in the current context. For our purposes it suffices to restrict our analysis here to the upstairs GUT vector-like spectrum.

Since the multiplicity operator~\eqref{MultiOp4D} is insensitive to fully vector-like states, we calculate their spectrum via equivariant line bundle cohomology.\footnote{Computation of the full spectrum is unfeasible for the Schoen manifold due to its large Stanley--Reissner ideal. Hence we omit it from our discussion here.} For this we use the automated computer tool \texttt{cohomcalg} \cite{Blumenhagen:2010pv,cohomCalg:Implementation}. Starting from those models that are chiral exact, we compute the full spectrum to see whether they have at least one non-chiral  $\rep{5}$-$\crep{5}$ pair. 
(In a few cases the exact number of non-chiral pairs is not fixed uniquely and depends on further input data.) 
From such a pair we could obtain Higgs doublets without their triplet partners via a suitable choice of a Wilson line. An additional source for vector-like exotics could be pairs of $\rep{10}$-$\brep{10}$-plets. They will also split under the action of the freely acting Wilson line and thus vector-like pairs of exotics can remain in the downstairs spectrum. 

In Table~\ref{tab:FullSpectrumAnalysis} we collect the results of the analysis of the full spectrum for our smooth chiral exact models discussed in the previous section. When the chiral scans of the previous subsection did not find any chiral exact models on a given geometry, we ignore this geometry in the current analysis; for this reason e.g.\ CICY 6225 is absent in Table~\ref{tab:FullSpectrumAnalysis}. In the second-to-last column we indicate the percentage of those models which have at least one $\rep{5}$-$\crep{5}$ Higgs pair candidate and no ($\rep{10}$-$\crep{10}$) exotics. These percentages are defined w.r.t.\ the total number of chiral exact E$_8\times$E$_8$ and SO(32) models given in Tables~\ref{tb:ScanOverviewE8xE8} and~\ref{tb:ScanOverviewSO32}. In the last column  we list models which, apart from having at least one Higgs pair candidate, possess additional vector-like exotic states. These exotics can be additional $\rep{5}$-$\crep{5}$ and $\rep{10}$-$\crep{10}$ pairs. If the numbers do not sum up to 100\%, the rest of the models do not have a $\rep{5}$-$\brep{5}$ pair and hence no Higgs candidate. For these we do not distinguish whether or not there are additional exotic $\rep{10}$-$\crep{10}$ pairs present.

We observe that most of the chiral exact MSSM models of the supersymmetric theories do have vector-like $\rep{5}$-$\brep{5}$ pairs which may serve as Higgs candidates in the downstairs spectrum. We also find that the number of vector-like exotics (beyond multiple Higgses) is quite small or even negligible in almost all cases. Moreover, such vector-like exotics are more abundant in the E$_8\times$E$_8$ theory than in the SO(32) theory.

We have not included any GUT-like models from the non-supersymmetric SO(16)$\times$SO(16) theory in Table~\ref{tab:FullSpectrumAnalysis} since for non-supersymmetric models the definition of chiral exactness differs from that of the supersymmetric ones in that chiral scalars are distinguished from fermionic exotics. Nevertheless, we have analyzed the spectra of the six models that are both fermionic and bosonic chirally exact on the two CICY geometries 7487 and 7862 and found that each of these models contains three scalar five-plet vector-like pairs, i.e.\ Higgs boson candidates, and no additional vector-like Standard Model matter.

\subsection{Distribution of the number of chiral generations}\label{sc:SU5_statistics}

\begin{figure}
\begin{center}  
\includegraphics[scale=0.6]{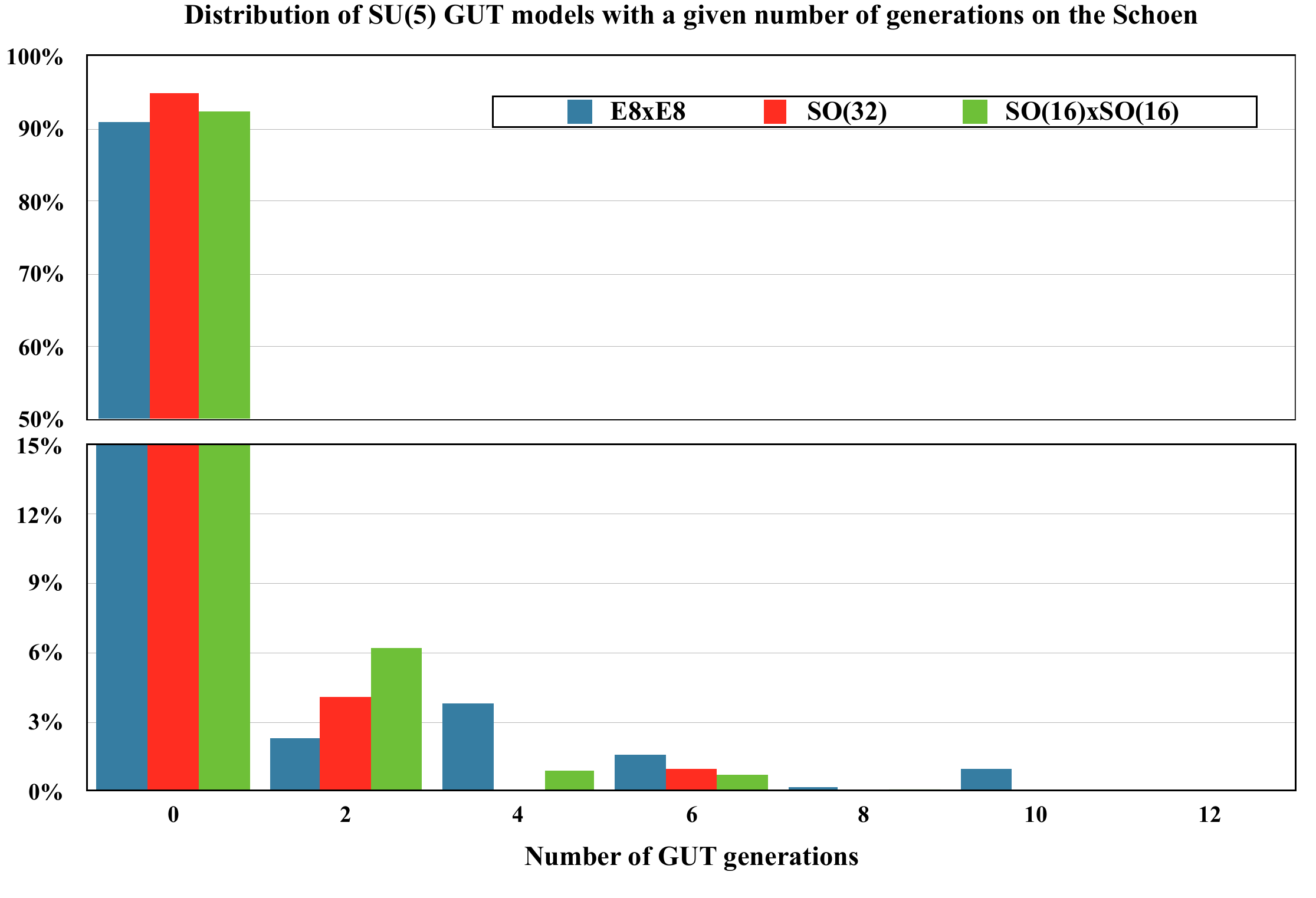}
\includegraphics[scale=0.606]{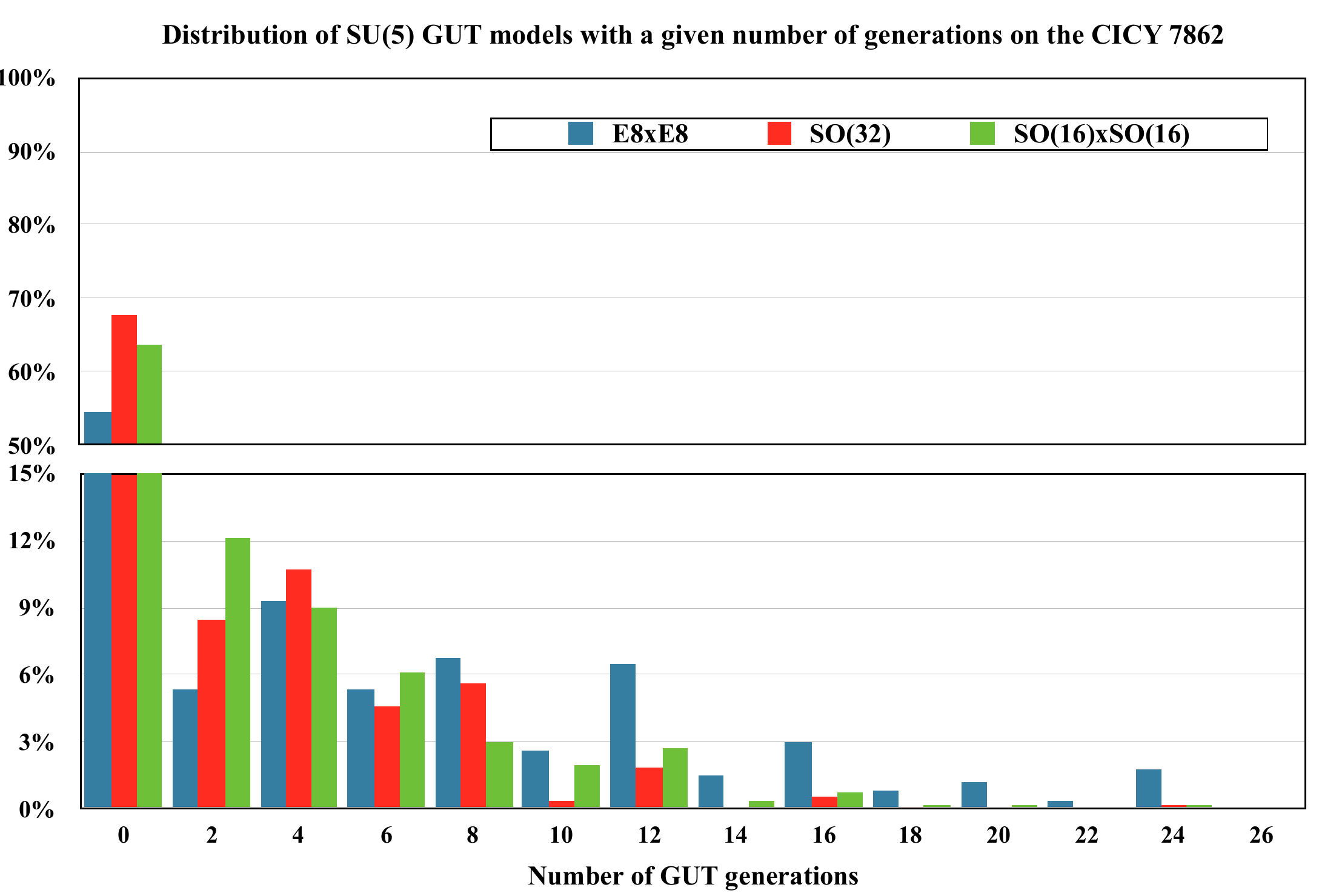}
\end{center}
\caption{\label{fg:SU5GUT_Plots}
The two plots show the portion of chiral SU(5)-GUT models with a given generation number (in the upstairs picture) for each of the three heterotic theories on the Schoen manifold and the tetra-quadric, normalized w.r.t.\ the total number of SU(5) models we have constructed (well over $10^7$ inequivalent SU(5) models for each theory on the Schoen manifold and well over $10^6$ for each theory on the tetra-quadric).
}
\end{figure}

In this subsection we study the likelihood that a given GUT model leads to three generations in the downstairs picture using the geometries and the line bundle construction outlined in the previous sections. To answer this question statistically, we have plotted the percentage of SU(5) GUT-like models with $N$ GUT generations against the total number of models with SU(5) gauge group. For this discussion all figures refer to inequivalent models as defined in Section~\ref{sc:WorkingDefinitions}. In the two histograms of Figure~\ref{fg:SU5GUT_Plots} we present our statistical analysis for the Schoen manifold and the tetra-quadric, respectively, for all three heterotic theories side-by-side. Let us emphasize a few interesting features of these plots: 

First, we see that on both geometries the majority (in most cases more than 70\%) of SU(5) GUT models are non-chiral, i.e.\ they have zero net number of GUT generations. The peaks for the number of generations lies at either 2 or 4, depending on the geometry and theory under consideration. The distribution dies off rather slowly especially for the E$_8\times$E$_8$ on the tetra-quadric. 
In other words we see that the number of GUT generations is scattered over a much larger range for the favorable CICYs with small $h_{11}$ than on the Schoen manifold, which has a clear preference for a small number of GUT-generations (up to ten at most). For these reasons we have plotted these distributions in Figure~\ref{fg:SU5GUT_Plots} with two different scales to visualize both the peak of non-chiral models and how the net chiral GUT models are distributed. 

For both geometries we mod out an order 2 Wilson line, hence six net GUT generations would be phenomenologically preferred.
However, we see that this only happens in at most 2\% of the constructed SU(5) models on the Schoen manifold and at most 6\% of the cases on the tetra-quadric. Hence, these statistics do not seem to indicate that (MS)SMs with three generations are singled out or favored in any way.
Instead, we have to veto thousands of phenomenologically uninteresting models, before actually obtaining any relevant GUT theory.

\begin{table}[t]
\begin{center}
\renewcommand{\arraystretch}{1.2}
\begin{tabular}{|c||c||c|}
\hline 
\multicolumn{3}{|c|}{\bf SO(32) MSSM on CICY 7862  with line bundle vectors and $\boldsymbol{\Intr_2}$ Wilson line:}
\\ \hline\hline
\multicolumn{3}{|l|}{
\scalebox{.9}{\(
\begin{array}{lcrrrrrrrrrrrrrrrrrl}
V_1 & = & ( &    -2 , &    -2 , &    -2 , &    -2 , &    -2 , &     0 , &    -1 , &    -3 , &     0 , &     1 , &     0 , &    -1 , &    -1 , &     1 , &     1 , &    -1 & )
\\
V_2 & = & ( &     0 , &     0 , &     0 , &     0 , &     0 , &     2 , &    -1 , &     1 , &    -2 , &    -3 , &    -2 , &    -1 , &    -1 , &     1 , &     1 , &     3 & )
\\
V_3 & = & ( &     0 , &     0 , &     0 , &     0 , &     0 , &     0 , &     0 , &    -1 , &     0 , &     1 , &     1 , &     1 , &     0 , &     0 , &    -1 , &    -1 & )
\\
V_4 & = & ( & \tfrac{1}{2} , & \tfrac{1}{2} , & \tfrac{1}{2} , & \tfrac{1}{2} , & \tfrac{1}{2} , & -\tfrac{1}{2} , & \tfrac{1}{2} , & \tfrac{3}{2} , & \tfrac{1}{2} , & -\tfrac{1}{2} , & -\tfrac{1}{2} , & -\tfrac{1}{2} , & \tfrac{1}{2} , & -\tfrac{1}{2} , & \tfrac{1}{2} , & \tfrac{1}{2} & )
\\[1ex] 
W & = & ( & \tfrac{1}{2} , & \tfrac{1}{2} , & \tfrac{1}{2} , &     0 , &     0 , & \tfrac{1}{2} , &     1 , & -\tfrac{3}{2} , &     0 , &     0 , &     0 , &     0 , & \tfrac{1}{2} , &     0 , & \tfrac{1}{2} , & \tfrac{1}{2} & )
\end{array}
\)}
}
\\[-3ex] \multicolumn{3}{|l|}{}
\\ \hline\hline 
& {\bf Upstairs spectrum}	&  {\bf Downstairs spectrum} 
\\
& 	
SU(5) $\times$ SU(3)$\times$SU(2)$^{3}$		
&	
SU(3)$\times$SU(2)$\times$U(1)$_Y$ $\times$ SU(2)
\\ \hline\hline 
\multirow{4}{*}{\rotatebox{90}{\bf observable}}
&		& \\[-2ex] 	
& 
$6\, (\textbf{10};  \textbf{1},  \textbf{1},  \textbf{1},  \textbf{1})$ 		
& 
$3\, (\textbf{3},\textbf{2};\textbf{1})_{-\sfrac16} + 3\, (\overline{\textbf{3}},\textbf{1};\textbf{1})_{\sfrac23} + 3\, (\textbf{1},\textbf{1};\textbf{1})_{-1}$
\\
&		& \\[-2ex] 			
& 
$2\, (\overline{\textbf{5}};  \textbf{1},  \textbf{1},  \textbf{1},  \textbf{1}) + 2\, (\overline{\textbf{5}};  \textbf{1},  \textbf{2},  \textbf{1},  \textbf{1})$ 	
& 
$3\, (\overline{\textbf{3}},\textbf{1};\textbf{1})_{-\sfrac13} + 3\, (\textbf{1},\textbf{2};\textbf{1})_{\sfrac12}$
\\[-2ex] 
&                   & 
\\ \hline\hline
\multirow{5}{*}{\rotatebox{90}{\bf hidden}}	& 
$10\, (  \textbf{1}; \overline{\textbf{3}},  \textbf{1},  \textbf{1},  \textbf{1}) + 2\, (  \textbf{1};  \textbf{3},  \textbf{1},  \textbf{1},  \textbf{1}) +$                             
&  
\multirow{2}{*}{$14\, (\textbf{1},\textbf{1};\textbf{2})_{0} + 14\, (\textbf{1},\textbf{1};\textbf{1})_{0}$} \\
&  
$6\, (  \textbf{1};  \textbf{3},  \textbf{1},  \textbf{2},  \textbf{1}) + 2\, (  \textbf{1}; \overline{\textbf{3}},  \textbf{1},  \textbf{1},  \textbf{2})$                        &  \\[1ex]
&  
$16\, (  \textbf{1};  \textbf{1},  \textbf{2},  \textbf{2},  \textbf{1})+ 10\, (  \textbf{1};  \textbf{1},  \textbf{1},  \textbf{2},  \textbf{2}) + $                           
&   \\
&  
$22\, (  \textbf{1};  \textbf{1},  \textbf{1},  \textbf{1},  \textbf{2}) + 22\, (  \textbf{1};  \textbf{1},  \textbf{2},  \textbf{1},  \textbf{1}) +$                           
& 
$117\, (\textbf{1},\textbf{1};\textbf{1})_{0}$\\
&  
$2\, (  \textbf{1};  \textbf{1},  \textbf{1},  \textbf{2},  \textbf{1}) + 38\, (  \textbf{1};  \textbf{1},  \textbf{1},  \textbf{1},  \textbf{1})$                           
&  \\
\hline
\end{tabular}
\renewcommand{\arraystretch}{1}
\end{center}
\caption{\label{tb:SO32_MSSM}
A chiral exact MSSM-like model from the supersymmetric SO(32) theory on the tetra-quadric. The left part of the table gives the upstairs GUT-like spectrum and the right part the resulting downstairs spectrum after Wilson line symmetry breaking. 
For clarity we only display the hypercharge and omit the other U(1) charges.
}
\end{table}


\section{Examples of (MS)SM-like models}

In this section we present explicit examples of an MSSM-like model from the SO(32) theory and an SM-like model from the SO(16)$\times$SO(16) theory, both on the tetra-quadric (CICY 7862) geometry. These examples are chiral exact models, the first in the supersymmetric sense defined above and the latter at the level of fermions. Since for the E$_8\times$E$_8$ theory an example of a chiral exact MSSM-like model was presented in~\cite{Nibbelink:2015ixa} with an extended analysis of its EFT, we refer to that paper for a comprehensive presentation for a chiral exact example from the E$_8\times$E$_8$ theory.

\subsubsection*{A chiral exact MSSM-like heterotic SO(32) model}

In Table~\ref{tb:SO32_MSSM} we consider a specific example for an SO(32) smooth model. We present the defining line bundle vectors and the $\Intr_2$ Wilson line, as well as the resulting upstairs and downstairs spectra for the hidden and the observable sector. The rows are organized such that the projection and branching due to the Wilson line can be reconstructed for the various upstairs to downstairs representations. One may verify that all gauge anomalies are cancelled. 

Note that the number of $\crep{3}$- and $\rep{2}$-plets in the downstairs spectrum arise from two types of states in the upstairs picture: 
$(\overline{\textbf{5}};  \textbf{1},  \textbf{1},  \textbf{1},  \textbf{1})$ and $(\overline{\textbf{5}};  \textbf{1},  \textbf{2},  \textbf{1},  \textbf{1})$ both of which have multiplicity 2. The downstairs multiplicity is obtained as: $(2 + 2 \cdot 2)/2 = 3$. The additional SU(2) under which the $(\overline{\textbf{5}};  \textbf{1},  \textbf{2},  \textbf{1},  \textbf{1})$ are charged may therefore be a way to distinguish the first two lighter generations from the third. 

\subsubsection*{A chiral exact SM-like heterotic SO(16)$\boldsymbol{\times}$SO(16) model}

Next, we present one of the SO$(16)\times$SO$(16)$ fermionic chirally exact SM-like models. The model is summarized in Table~\ref{tb:SO16_SM}. Since the model is non-supersymmetric we list the complex bosons and chiral fermions separately on gray and white backgrounds, respectively. It might seem that this model contains many more states than the supersymmetric examples. However, one should keep in mind that we give the spectra of bosons and fermions separately for the non-supersymmetric GUT models. In contrast, for the supersymmetric models we follow the standard convention to give the spectra in terms of superfields. If one were to write out the full bosonic and fermionic spectra of the superfields, one would find that the non-supersymmetric models contain less scalars than their supersymmetric partners. In fact, this is to be expected, since the supersymmetric theories in ten dimensions contain 496 gauge fields and gauginos, while the non-supersymmetric SO(16)$\times$SO(16) has only 240 gauge fields and 512 charged fermions.

It is amusing to note that exactly the same bundle and Wilson line provide us with an MSSM-like model for the E$_8\times$E$_8$ theory; however this MSSM-like model is not chiral exact.  
The feature that the same bundle leads to both MSSM- and SM-like models is not actually that rare and demonstrates again how closely related the theories are, at least to leading order at the level of massless spectrum. 

\clearpage 

\begin{table}[!t]
\begin{center}
\renewcommand{\arraystretch}{1.2}
\begin{tabular}{|c||c||c|}
\hline 
\multicolumn{3}{|c|}{\bf SO(16)$\boldsymbol{\times}$SO(16) SM on CICY 7862  with line bundle vectors and $\boldsymbol{\Intr_2}$ Wilson line:}
\\ \hline\hline
\multicolumn{3}{|l|}{
\scalebox{.9}{\(
\begin{array}{lcrrrrrrrrrrrrrrrrrrl}
V_1 & = & ( &     1 , &    -1 , &    -1 , &    -1 , &     1 , &     3 , &     1 , &     1 &)  ( &    0 , &     0 , &     0 , &     0 , &     0 , &     0 , &     0 , &     0 & )
\\
V_2 & = & ( &     0 , &     0 , &     0 , &     0 , &     0 , &    -1 , &     0 , &    -1 &)  ( &  1 , &     0 , &    -1 , &     1 , &     1 , &     0 , &     0 , &     0 & )
\\
V_3 & = & ( &    -1 , &     1 , &     1 , &     1 , &    -1 , &     0 , &    -2 , &    -1 &)  ( & -1 , &     0 , &     0 , &     0 , &     0 , &     0 , &    -1 , &     0 & )
\\
V_4 & = & ( &     0 , &     0 , &     0 , &     0 , &     0 , &    -1 , &     1 , &     2 &)  ( &  -1 , &     0 , &     2 , &    -2 , &    -2 , &     0 , &     1 , &     0 & )
\\[1ex] 
W & = & ( & -\tfrac{7}{4} , & -\tfrac{1}{4} , & -\tfrac{1}{4} , & \tfrac{1}{4} , & \tfrac{7}{4} , & -\tfrac{1}{4} , & \tfrac{1}{4} , & \tfrac{1}{4} &)  ( & -\tfrac{1}{4} , & -\tfrac{1}{4} , & -\tfrac{1}{4} , & -\tfrac{1}{4} , & \tfrac{1}{4} , & \tfrac{1}{4} , & \tfrac{1}{4} , & \tfrac{1}{4} & )
\end{array}
\)}
}
\\[-3ex] \multicolumn{3}{|l|}{}
\\ \hline\hline 
&	{SU$(5)'\times$SU$(4)''\times$SU$(3)''$}	&	{SU$(3)'\times$SU$(2)'\times$SU$(3)''\times$SU$(2)''\times$U$(1)'_Y$}
\\\hline\hline 
\multirow{4}{*}{\rotatebox{90}{observable}}	
&	 \cellcolor{lightgray}$24\, (\overline{\textbf{5}};  \textbf{1},  \textbf{1})$ 	
& 	\cellcolor{lightgray}$12\, (\overline{\textbf{3}},\textbf{1};\textbf{1},\textbf{1})_{-\sfrac13} + 12\, (\textbf{1},\textbf{2};\textbf{1},\textbf{1})_{\sfrac12}$ 
\\
&	\cellcolor{lightgray}$20\, (\textbf{5}; \textbf{1}, \textbf{1})$ 					
&	\cellcolor{lightgray}$10\, (\textbf{3},\textbf{1};\textbf{1},\textbf{1})_{\sfrac13} + 10\,(\textbf{1},\textbf{2};\textbf{1},\textbf{1})_{-\sfrac12}$ 
\\[1ex] \cline{2-3} 
&	$6\,(\textbf{10};  \textbf{1},  \textbf{1})$ 		
&	$3\,(\textbf{3},\textbf{2};\textbf{1},\textbf{1})_{-\sfrac16} + 3\, (\overline{\textbf{3}},\textbf{1};\textbf{1},\textbf{1})_{\sfrac23} + 3\, (\textbf{1},\textbf{1};\textbf{1},\textbf{1})_{-1}$
\\			
&	$6\, (\overline{\textbf{5}};  \textbf{1})$ 		    
&   $3\, (\overline{\textbf{3}},\textbf{1};\textbf{1},\textbf{1})_{-\sfrac13} + 3\, (\textbf{1},\textbf{2};\textbf{1},\textbf{1})_{\sfrac12}$ 
\\[-3ex] && 
\\ \hline\hline
\multirow{10}{*}{\rotatebox{90}{hidden}}	
&	\cellcolor{lightgray}$2\,(\textbf{1};  \textbf{6}, \overline{\textbf{3}})$ 	
&	\cellcolor{lightgray}$(\textbf{1},\textbf{1};\textbf{3},\textbf{2})_{0} + (\textbf{1},\textbf{1};\overline{\textbf{3}},\textbf{2})_{0} + (\textbf{1},\textbf{1};\overline{\textbf{3}},\textbf{1})_{0} + (\textbf{1},\textbf{1};\textbf{3},\textbf{1})_{0}$ 
\\
&	\cellcolor{lightgray} $12\, (\textbf{1}; \textbf{1}, \textbf{3})$ 	
&	\cellcolor{lightgray} $6\,(\textbf{1},\textbf{1};\textbf{1},\textbf{2})_{0}+ 6\,(\textbf{1},\textbf{1};\textbf{1},\textbf{1})_{0}$ 
\\
&	\cellcolor{lightgray}$8\, (\textbf{1};  \textbf{1}, \overline{\textbf{3}})$ 
&	\cellcolor{lightgray}$4\,(\textbf{1},\textbf{1};\textbf{1},\textbf{2})_{0}+ 4\,(\textbf{1},\textbf{1};\textbf{1},\textbf{1})_{0}$ 
\\
&	\cellcolor{lightgray}$56\, (\textbf{1}; \textbf{1}, \textbf{1})$ 
&	\cellcolor{lightgray}$28\,(\textbf{1},\textbf{1};\textbf{1},\textbf{1})_{0}$ 
\\
\cline{2-3}
&	$2\, (  \textbf{1};  \overline{\textbf{4}},  \textbf{3})$                    
&	$(\textbf{1},\textbf{1};\overline{\textbf{3}},\textbf{2})_{0}+ (\textbf{1},\textbf{1};\textbf{1},\textbf{2})_{0}+ (\textbf{1},\textbf{1};\overline{\textbf{3}},\textbf{1})_{0}+(\textbf{1},\textbf{1};\textbf{1},\textbf{1})_{0}$ 
\\
&	$12\, (  \textbf{1};  \textbf{6},  \textbf{1})$                     
&	$6\,(\textbf{1},\textbf{1};\textbf{3},\textbf{1})_{0} + 6\,(\textbf{1},\textbf{1};\overline{\textbf{3}},\textbf{1})_{0}$ 
\\
&	$14\, (  \textbf{1};  \textbf{4},  \textbf{1})$     
&	$7\,(\textbf{1},\textbf{1};\textbf{3},\textbf{1})_{0}+ 7\,(\textbf{1},\textbf{1};\textbf{1},\textbf{1})_{0}$ 
\\
&	 $8\,(  \textbf{1};  \overline{\textbf{4}},  \textbf{1})$     
&	$4\,(\textbf{1},\textbf{1};\overline{\textbf{3}},\textbf{1})_{0} + 4\,(\textbf{1},\textbf{1};\textbf{1},\textbf{1})_{0}$ 
\\
&	$8\,(  \textbf{1};  \textbf{1},  \overline{\textbf{3}})$          
&	$4\, (\textbf{1},\textbf{1};\textbf{1},\textbf{2})_{0}+ 4\,(\textbf{1},\textbf{1};\textbf{1},\textbf{1})_{0}$ 
\\
&	$134\, (  \textbf{1};  \textbf{1},  \textbf{1})$                        
&	$67\,(\textbf{1},\textbf{1};\textbf{1},\textbf{1})_{0}$ 
\\ \hline
\end{tabular}
\renewcommand{\arraystretch}{1}
\end{center}
\caption{\label{tb:SO16_SM}
A chiral exact SM-like model from the SO$(16)\times$SO$(16)$ theory on the tetra-quadric. In this table we use the same conventions as in Table~\ref{tb:SO32_MSSM}, except that we depict the complex bosons and chiral fermions on gray and white backgrounds, respectively. 
}
\end{table}

\clearpage

\newpage

\section{Conclusions}
\label{sc:Conclusion}

We have performed model searches on smooth Calabi-Yau compactifications for both supersymmetric heterotic E$_8\times$E$_8$ and SO(32) theories and for the non-supersymmetric SO(16)$\times$SO(16) theory simultaneously. 
As far as we are aware this is the first work in which all these three theories are compared in this side-by-side fashion. 

Considering smooth Calabi-Yau compactifications of all three heterotic theories is particularly beneficial for the non-supersymmetric SO(16)$\times$SO(16) theory, since it is guaranteed to avoid tachyons at leading order in $\ga'$ and $g_s$. Furthermore, we can make use of many known methods to analyze the phenomenological properties of such constructions. We focused on a subset of favorable CICYs of relatively small $h_{11} \leq 6$. In addition, we considered the Schoen manifold, described as an orbifold resolution, to have an example with larger $h_{11} (=19)$. 
In order to obtain some systematic results we considered exclusively line bundle gauge embedding on these geometries. 

Generic line bundle backgrounds can be characterized by a collection of $h_{11}$ sixteen-component bundle vectors $V_i$. 
Using this description the Bianchi identities without NS5-branes are quadratic Diophantine equations which are very hard to solve. 
For this reason we introduced Gram matrices, defined as $K_{ij} = V_i \cdot V_j$, to systematically analyze the combined consequences of the Bianchi identities and the tree-level DUY equations. 
If one takes a point inside the K\"ahler cone as input, these equations can be interpreted as a linear system of $2\cdot h_{11}$ equations of $\sfrac 12\, h_{11}(h_{11}+1)$ parameters. 
Not all such solutions can be used for model building as not all of them can be written as Gram matrices. For those that can, we generate corresponding sets of line bundle vectors for the three heterotic theories.  
Since the lattice on which the bundle vectors of the non-supersymmetric theory lie is a sublattice of both the E$_8\times$E$_8$ and SO(32) lattices, all bundle vector solutions of the SO(16)$\times$SO(16) theory are automatically also solutions of the supersymmetric theories.

We construct the full charged chiral spectrum of these models by exploiting the multiplicity operator in four dimensions. 
For each of the spectra we check all pure non-Abelian as well as mixed Abelian-gravitational and pure Abelian anomalies using the generalized Green-Schwarz mechanism with $h_{11}+1$ universal and non-universal axion couplings. 
This constitutes many consistency checks on the charged chiral spectrum. 
(For example, for $h_{11}=5$ and only two non-Abelian SU($N$) factors we already have more than 50 anomaly conditions.) 

In order to have a way of comparing the model building potential of the various heterotic theories on the various geometries, we have preformed computer-aided scans for all cases for a fixed period of time. 
The total model search duration seems to be arbitrarily chosen; indeed, as was pointed out in~\cite{Nibbelink:2015ixa}, there does not seem to be a clear-cut bound on the line bundle input data. Consequently, we arbitrarily constrained the duration of scans to the same period of time for each heterotic theory on the various geometries under consideration.

For all three heterotic string theories we have generated a large number of GUT-like models (up to over a few hundred thousand) which become (up to a few million) (MS)SM-like models upon using a freely acting Wilson line. 
We find that having three generation models does not seem to be especially singled out. 
Chiral exact models are not as abundant as models with additional vector-like SM-fermions.
However, we were able to construct chiral exact non-supersymmetric SM-like models, in both the bosonic and fermionic sense, which even contain Higgs candidates in the form of vector-like five-plet pairs.

\subsection*{Acknowledgements}

We thank Michael Blaszczyk, Steve Abel, Saul Ramos-Sanchez and Patrick K.S.~Vaudrevange for valuable discussions about (non-)supersymmetric model building. 

The work of F.R.\ was supported by the German Science Foundation (DFG) within the Collaborative Research Center (SFB) 676 ``Particles, Strings and the Early Universe''. 
O.L. acknowledges the support by the DAAD Scholarship Programme ``Vollstipendium f\"{u}r Absolventen von deutschen Auslandsschulen'' within the ``PASCH--Initiative''.

{\small
\providecommand{\href}[2]{#2}\begingroup\raggedright\endgroup

}

\begin{thebibliography}{10}

\bibitem{Nibbelink:2009sp}
S.~Groot~Nibbelink, J.~Held, F.~Ruehle, M.~Trapletti, and P.~K.~S. Vaudrevange
  ``{Heterotic Z6-II MSSM orbifolds in blowup}'' {\em JHEP} {\bf 03} (2009) 005
\href{http://www.arXiv.org/abs/0901.3059}{[{\tt arXiv:0901.3059}]}.

\bibitem{Blaszczyk:2010db}
M.~Blaszczyk, S.~Groot~Nibbelink, F.~Ruehle, M.~Trapletti, and P.~K.~S.
  Vaudrevange ``{Heterotic MSSM on a resolved orbifold}'' {\em JHEP} {\bf 09}
  (2010) 065
\href{http://www.arXiv.org/abs/1007.0203}{[{\tt arXiv:1007.0203}]}.

\bibitem{Nibbelink:2012de}
S.~Groot~Nibbelink and P.~K. Vaudrevange ``{Schoen manifold with line bundles
  as resolved magnetized orbifolds}'' {\em JHEP} {\bf 1303} (2013) 142
\href{http://www.arXiv.org/abs/1212.4033}{[{\tt arXiv:1212.4033}]}.

\bibitem{Anderson:2011ns}
L.~B. Anderson, J.~Gray, A.~Lukas, and E.~Palti ``{Two hundred heterotic
  Standard Models on smooth Calabi-Yau threefolds}''
  \href{http://www.arXiv.org/abs/1106.4804}{[{\tt arXiv:1106.4804}]}.

\bibitem{Anderson:2012yf}
L.~B. Anderson, J.~Gray, A.~Lukas, and E.~Palti ``{Heterotic line bundle
  Standard Models}'' {\em JHEP} {\bf 1206} (2012) 113
\href{http://www.arXiv.org/abs/1202.1757}{[{\tt arXiv:1202.1757}]}.

\bibitem{Anderson:2013xka}
L.~B. Anderson, A.~Constantin, J.~Gray, A.~Lukas, and E.~Palti ``{A
  comprehensive scan for heterotic SU(5) GUT models}'' {\em JHEP} {\bf 1401}
  (2014) 047
\href{http://www.arXiv.org/abs/1307.4787}{[{\tt arXiv:1307.4787}]}.

\bibitem{Blumenhagen:2005zg}
R.~Blumenhagen, G.~Honecker, and T.~Weigand ``{Non-Abelian brane worlds: The
  heterotic string story}'' {\em JHEP} {\bf 10} (2005) 086
\href{http://www.arXiv.org/abs/hep-th/0510049}{[{\tt arXiv:hep-th/0510049}]}.

\bibitem{Font:2002pq}
A.~Font and A.~Hern\'andez ``{Nonsupersymmetric orbifolds}'' {\em Nucl.Phys.}
  {\bf B634} (2002) 51--70
\href{http://www.arXiv.org/abs/hep-th/0202057}{[{\tt arXiv:hep-th/0202057}]}.

\bibitem{Blaszczyk:2014qoa}
M.~Blaszczyk, S.~Groot~Nibbelink, O.~Loukas, and S.~Ramos-Sanchez
  ``{Non-supersymmetric heterotic model building}'' {\em JHEP} {\bf 1410}
  (2014) 119
\href{http://www.arXiv.org/abs/1407.6362}{[{\tt arXiv:1407.6362}]}.

\bibitem{Blaszczyk:2015zta}
M.~Blaszczyk, S.~Groot~Nibbelink, O.~Loukas, and F.~Ruehle ``{Calabi-Yau
  compactifications of non-supersymmetric heterotic string theory}''
\href{http://www.arXiv.org/abs/1507.06147}{[{\tt arXiv:1507.06147}]}.

\bibitem{Buchmuller:2005jr}
W.~Buchm\"uller, K.~Hamaguchi, O.~Lebedev, and M.~Ratz ``{Supersymmetric
  Standard Model from the heterotic string}'' {\em Phys.Rev.Lett.} {\bf 96}
  (2006) 121602
\href{http://www.arXiv.org/abs/hep-ph/0511035}{[{\tt arXiv:hep-ph/0511035}]}.

\bibitem{Buchmuller:2006ik}
W.~Buchm{\"u}ller, K.~Hamaguchi, O.~Lebedev, and M.~Ratz ``{Supersymmetric
  Standard Model from the heterotic string}'' {\em Phys. Rev. Lett.} {\bf 96}
  (2006) 121602
\href{http://www.arXiv.org/abs/hep-ph/0511035}{[{\tt arXiv:hep-ph/0511035}]}.

\bibitem{Lebedev:2006kn}
O.~Lebedev {\em et al.} ``{A mini-landscape of exact MSSM spectra in heterotic
  orbifolds}'' {\em Phys. Lett.} {\bf B645} (2007) 88--94
\href{http://www.arXiv.org/abs/hep-th/0611095}{[{\tt arXiv:hep-th/0611095}]}.

\bibitem{Lebedev:2008un}
O.~Lebedev, H.~P. Nilles, S.~Ramos-S\'anchez, M.~Ratz, and P.~K. Vaudrevange
  ``{Heterotic mini-landscape. (II). Completing the search for MSSM vacua in a
  Z(6) orbifold}'' {\em Phys.Lett.} {\bf B668} (2008) 331--335
\href{http://www.arXiv.org/abs/0807.4384}{[{\tt arXiv:0807.4384}]}.

\bibitem{Z2xZ4}
D.~K. Mayorga~Pena, H.~P. Nilles, and P.-K. Oehlmann ``{A Zip-code for quarks,
  leptons and higgs bosons}'' {\em JHEP} {\bf 1212} (2012) 024
\href{http://www.arXiv.org/abs/1209.6041}{[{\tt arXiv:1209.6041}]}.

\bibitem{Kim:2006hv}
J.~E. Kim and B.~Kyae ``{String MSSM through flipped SU(5) from Z(12)
  orbifold}''
\href{http://www.arXiv.org/abs/hep-th/0608085}{[{\tt arXiv:hep-th/0608085}]}.

\bibitem{Kim:2007mt}
J.~E. Kim, J.-H. Kim, and B.~Kyae ``{Superstring Standard Model from Z(12-I)
  orbifold compactification with and without exotics, and effective R-parity}''
  {\em JHEP} {\bf 06} (2007) 034
\href{http://www.arXiv.org/abs/hep-ph/0702278}{[{\tt arXiv:hep-ph/0702278}]}.

\bibitem{Blaszczyk:2009in}
M.~Blaszczyk {\em et al.} ``{A Z2xZ2 standard model}'' {\em Phys. Lett.} {\bf
  B683} (2010) 340--348
\href{http://www.arXiv.org/abs/0911.4905}{[{\tt arXiv:0911.4905}]}.

\bibitem{Nibbelink:2013lua}
S.~Groot~Nibbelink and O.~Loukas ``{MSSM-like models on Z(8) toroidal
  orbifolds}'' {\em JHEP} {\bf 1312} (2013) 044
\href{http://www.arXiv.org/abs/1308.5145}{[{\tt arXiv:1308.5145}]}.

\bibitem{Nilles:2014owa}
H.~P. Nilles and P.~K.~S. Vaudrevange ``{Geography of fields in extra
  dimensions: String theory lessons for particle physics}''
\href{http://www.arXiv.org/abs/1403.1597}{[{\tt arXiv:1403.1597}]}.

\bibitem{Cleaver:1998sa}
G.~B. Cleaver, A.~E. Faraggi, and D.~V. Nanopoulos ``String derived {MSSM} and
  {M}-theory unification'' {\em Phys. Lett.} {\bf B455} (1999) 135--146
\href{http://www.arXiv.org/abs/hep-ph/9811427}{[{\tt arXiv:hep-ph/9811427}]}.

\bibitem{Faraggi:2007tj}
A.~E. Faraggi and M.~Tsulaia ``{On the low energy spectra of the
  nonsupersymmetric heterotic string theories}'' {\em Eur.Phys.J.} {\bf C54}
  (2008) 495--500
\href{http://www.arXiv.org/abs/0706.1649}{[{\tt arXiv:0706.1649}]}.

\bibitem{Faraggi:2014hqa}
A.~Faraggi, J.~Rizos, and H.~Sonmez ``{Classification of flipped SU(5)
  heterotic-string vacua}'' {\em Nucl. Phys.} {\bf B886} (2014) 202--242
\href{http://www.arXiv.org/abs/1403.4107}{[{\tt arXiv:1403.4107}]}.

\bibitem{Faraggi:2014vma}
A.~E. Faraggi and H.~Sonmez ``{Classification of SU(4) X SU(2) X U(1)
  heterotic-string models}'' {\em Phys. Rev.} {\bf D91} (2015) 066006
\href{http://www.arXiv.org/abs/1412.2839}{[{\tt arXiv:1412.2839}]}.

\bibitem{Dijkstra:2004ym}
T.~P.~T. Dijkstra, L.~R. Huiszoon, and A.~N. Schellekens ``{Chiral
  supersymmetric standard model spectra from orientifolds of Gepner models}''
  {\em Phys. Lett.} {\bf B609} (2005) 408--417
\href{http://www.arXiv.org/abs/hep-th/0403196}{[{\tt arXiv:hep-th/0403196}]}.

\bibitem{Dijkstra:2004cc}
T.~P.~T. Dijkstra, L.~R. Huiszoon, and A.~N. Schellekens ``{Supersymmetric
  Standard Model spectra from RCFT orientifolds}'' {\em Nucl. Phys.} {\bf B710}
  (2005) 3--57
\href{http://www.arXiv.org/abs/hep-th/0411129}{[{\tt arXiv:hep-th/0411129}]}.

\bibitem{Rohm:1983aq}
R.~Rohm ``{Spontaneous supersymmetry breaking in supersymmetric string
  theories}'' {\em Nucl.Phys.} {\bf B237} (1984)
553.

\bibitem{Kounnas:1989dk}
C.~Kounnas and B.~Rostand ``{Coordinate Dependent Compactifications and
  Discrete Symmetries}'' {\em Nucl. Phys.} {\bf B341} (1990)
641--665.

\bibitem{Itoyama:1986ei}
H.~Itoyama and T.~R. Taylor ``{Supersymmetry restoration in the compactified
  O(16) x O(16)-prime heterotic string theory}'' {\em Phys. Lett.} {\bf B186}
  (1987)
129.

\bibitem{Kutasov:1990sv}
D.~Kutasov and N.~Seiberg ``{Number of degrees of freedom, density of states
  and tachyons in string theory and CFT}'' {\em Nucl. Phys.} {\bf B358} (1991)
600--618.

\bibitem{Dienes:1994np}
K.~R. Dienes ``{Modular invariance, finiteness, and misaligned supersymmetry:
  New constraints on the numbers of physical string states}'' {\em Nucl.Phys.}
  {\bf B429} (1994) 533--588
\href{http://www.arXiv.org/abs/hep-th/9402006}{[{\tt arXiv:hep-th/9402006}]}.

\bibitem{Dienes:2006ut}
K.~R. Dienes ``{Statistics on the heterotic landscape: Gauge groups and
  cosmological constants of four-dimensional heterotic strings}'' {\em
  Phys.Rev.} {\bf D73} (2006) 106010
\href{http://www.arXiv.org/abs/hep-th/0602286}{[{\tt arXiv:hep-th/0602286}]}.

\bibitem{Angelantonj:2010ic}
C.~Angelantonj, M.~Cardella, S.~Elitzur, and E.~Rabinovici ``{Vacuum stability,
  string density of states and the Riemann zeta function}'' {\em JHEP} {\bf 02}
  (2011) 024
\href{http://www.arXiv.org/abs/1012.5091}{[{\tt arXiv:1012.5091}]}.

\bibitem{Toon:1990ij}
A.~Toon ``{Nonsupersymmetric Z(4) orbifolds and Atkin-Lehner symmetry}'' {\em
  Phys.Lett.} {\bf B243} (1990)
68--72.

\bibitem{Harvey:1998rc}
J.~A. Harvey ``{String duality and nonsupersymmetric strings}'' {\em Phys.
  Rev.} {\bf D59} (1999) 026002
\href{http://www.arXiv.org/abs/hep-th/9807213}{[{\tt arXiv:hep-th/9807213}]}.

\bibitem{Aldazabal:1999jr}
G.~Aldazabal and A.~M. Uranga ``{Tachyon free nonsupersymmetric type IIB
  orientifolds via Brane - anti-brane systems}'' {\em JHEP} {\bf 10} (1999) 024
\href{http://www.arXiv.org/abs/hep-th/9908072}{[{\tt arXiv:hep-th/9908072}]}.

\bibitem{Antoniadis:1999xk}
I.~Antoniadis, E.~Dudas, and A.~Sagnotti ``{Brane supersymmetry breaking}''
  {\em Phys. Lett.} {\bf B464} (1999) 38--45
\href{http://www.arXiv.org/abs/hep-th/9908023}{[{\tt arXiv:hep-th/9908023}]}.

\bibitem{Angelantonj:1999gm}
C.~Angelantonj, I.~Antoniadis, and K.~Forger ``{Nonsupersymmetric type I
  strings with zero vacuum energy}'' {\em Nucl. Phys.} {\bf B555} (1999)
  116--134
\href{http://www.arXiv.org/abs/hep-th/9904092}{[{\tt arXiv:hep-th/9904092}]}.

\bibitem{Angelantonj:1999ms}
C.~Angelantonj, I.~Antoniadis, G.~D'Appollonio, E.~Dudas, and A.~Sagnotti
  ``{Type I vacua with brane supersymmetry breaking}'' {\em Nucl. Phys.} {\bf
  B572} (2000) 36--70
\href{http://www.arXiv.org/abs/hep-th/9911081}{[{\tt arXiv:hep-th/9911081}]}.

\bibitem{Dudas:2000ff}
E.~Dudas and J.~Mourad ``{Brane solutions in strings with broken supersymmetry
  and dilaton tadpoles}'' {\em Phys.Lett.} {\bf B486} (2000) 172--178
\href{http://www.arXiv.org/abs/hep-th/0004165}{[{\tt arXiv:hep-th/0004165}]}.

\bibitem{Dudas:2002dg}
E.~Dudas, J.~Mourad, and C.~Timirgaziu ``{Time and space dependent backgrounds
  from nonsupersymmetric strings}'' {\em Nucl.Phys.} {\bf B660} (2003) 3--24
\href{http://www.arXiv.org/abs/hep-th/0209176}{[{\tt arXiv:hep-th/0209176}]}.

\bibitem{Angelantonj:2007ts}
C.~Angelantonj and E.~Dudas ``{Metastable string vacua}'' {\em Phys.Lett.} {\bf
  B651} (2007) 239--245
\href{http://www.arXiv.org/abs/0704.2553}{[{\tt arXiv:0704.2553}]}.

\bibitem{GatoRivera:2007yi}
B.~Gato-Rivera and A.~Schellekens ``{Non-supersymmetric tachyon-free Type-II
  and Type-I closed strings from RCFT}'' {\em Phys.Lett.} {\bf B656} (2007)
  127--131
\href{http://www.arXiv.org/abs/0709.1426}{[{\tt arXiv:0709.1426}]}.

\bibitem{GatoRivera:2008zn}
B.~Gato-Rivera and A.~Schellekens ``{Non-supersymmetric orientifolds of Gepner
  models}'' {\em Phys.Lett.} {\bf B671} (2009) 105--110
\href{http://www.arXiv.org/abs/0810.2267}{[{\tt arXiv:0810.2267}]}.

\bibitem{Abel:2015oxa}
S.~Abel, K.~R. Dienes, and E.~Mavroudi ``{Towards a nonsupersymmetric string
  phenomenology}'' {\em Phys. Rev.} {\bf D91} (2015) no.~12, 126014
\href{http://www.arXiv.org/abs/1502.03087}{[{\tt arXiv:1502.03087}]}.

\bibitem{Ashfaque:2015vta}
J.~M. Ashfaque, P.~Athanasopoulos, A.~E. Faraggi, and H.~Sonmez
  ``{Non-tachyonic semi-realistic non-supersymmetric heterotic string vacua}''
\href{http://www.arXiv.org/abs/1506.03114}{[{\tt arXiv:1506.03114}]}.

\bibitem{Angelantonj:2014dia}
C.~Angelantonj, I.~Florakis, and M.~Tsulaia ``{Universality of Gauge Thresholds
  in Non-Supersymmetric Heterotic Vacua}'' {\em Phys. Lett.} {\bf B736} (2014)
  365--370
\href{http://www.arXiv.org/abs/1407.8023}{[{\tt arXiv:1407.8023}]}.

\bibitem{CYPage}
L.~B. Anderson, A.~Constantin, J.~Gray, A.~Lukas, and E.~Palti
  ``http://www-thphys.physics.ox.ac.uk/projects/calabiyau/linebundlemodels/index.html.''

\bibitem{Candelas:1987kf}
P.~Candelas, A.~Dale, C.~Lutken, and R.~Schimmrigk ``{Complete intersection
  Calabi-Yau manifolds}'' {\em Nucl.Phys.} {\bf B298} (1988)
493.

\bibitem{Braun:2010vc}
V.~Braun ``{O}n {F}ree {Q}uotients of {C}omplete {I}ntersection {C}alabi-{Y}au
  {M}anifolds'' {\em JHEP} {\bf 1104} (2011) 005
\href{http://www.arXiv.org/abs/1003.3235}{[{\tt arXiv:1003.3235}]}.

\bibitem{CYweb}
``{CALABI-YAU Home Page}'' 1996.
\newblock A resource for information about Calabi-Yau manifolds
  \href{http://www.th.physik.uni-bonn.de/Supplements/cy.html}{http://www.th.physik.uni-bonn.de/Supplements/cy.html}.

\bibitem{Hosono:1994ax}
S.~Hosono, A.~Klemm, S.~Theisen, and S.-T. Yau ``{M}irror symmetry, mirror map
  and applications to complete intersection {C}alabi-{Y}au spaces'' {\em
  Nucl.Phys.} {\bf B433} (1995) 501--554
\href{http://www.arXiv.org/abs/hep-th/9406055}{[{\tt arXiv:hep-th/9406055}]}.

\bibitem{Donagi:2008xy}
R.~Donagi and K.~Wendland ``{On orbifolds and free fermion constructions}''
  {\em J.Geom.Phys.} {\bf 59} (2009) 942--968
  \href{http://www.arXiv.org/abs/0809.0330}{[{\tt arXiv:0809.0330}]}.

\bibitem{Fischer:2012qj}
M.~Fischer, M.~Ratz, J.~Torrado, and P.~K. Vaudrevange ``{Classification of
  symmetric toroidal orbifolds}'' {\em JHEP} {\bf 1301} (2013) 084
\href{http://www.arXiv.org/abs/1209.3906}{[{\tt arXiv:1209.3906}]}.

\bibitem{Nibbelink:2007rd}
S.~{Groot Nibbelink}, M.~Trapletti, and M.~Walter ``Resolutions of {$C^n/Z_n$}
  orbifolds, their {U(1)} bundles, and applications to string model building''
  {\em JHEP} {\bf 03} (2007) 035
\href{http://www.arXiv.org/abs/hep-th/0701227}{[{\tt arXiv:hep-th/0701227}]}.

\bibitem{Nibbelink:2015ixa}
S.~Groot~Nibbelink, O.~Loukas, F.~Ruehle, and P.~K.~S. Vaudrevange ``{Infinite
  number of MSSMs from heterotic line bundles?}'' {\em To be published in Phys.
  Rev. D} (2015)
\href{http://www.arXiv.org/abs/1506.00879}{[{\tt arXiv:1506.00879}]}.

\bibitem{Blumenhagen:2005ga}
R.~Blumenhagen, G.~Honecker, and T.~Weigand ``Loop-corrected compactifications
  of the heterotic string with line bundles'' {\em JHEP} {\bf 06} (2005) 020
\href{http://www.arXiv.org/abs/hep-th/0504232}{[{\tt arXiv:hep-th/0504232}]}.

\bibitem{Blumenhagen:2006ux}
R.~Blumenhagen, S.~Moster, and T.~Weigand ``{Heterotic GUT and Standard Model
  vacua from simply connected Calabi-Yau manifolds}'' {\em Nucl. Phys.} {\bf
  B751} (2006) 186--221
\href{http://www.arXiv.org/abs/hep-th/0603015}{[{\tt arXiv:hep-th/0603015}]}.

\bibitem{Slansky:1981yr}
R.~Slansky ``{Group theory for unified model building}'' {\em Phys. Rept.} {\bf
  79} (1981)
1--128.

\bibitem{Nilles:2011aj}
H.~P. Nilles, S.~Ramos-S\'anchez, P.~K. Vaudrevange, and A.~Wingerter ``{The
  Orbifolder: A tool to study the low energy effective theory of heterotic
  orbifolds}'' {\em Comput.Phys.Commun.} {\bf 183} (2012) 1363--1380
\href{http://www.arXiv.org/abs/1110.5229}{[{\tt arXiv:1110.5229}]}.

\bibitem{Blumenhagen:2010pv}
R.~Blumenhagen, B.~Jurke, T.~Rahn, and H.~Roschy ``{Cohomology of line bundles:
  A computational algorithm}'' {\em J. Math. Phys.} {\bf 51} (2010) 103525
  \href{http://www.arXiv.org/abs/1003.5217}{[{\tt arXiv:1003.5217}]}.

\bibitem{cohomCalg:Implementation}
``{cohomCalg package}.'' Download link 2010.
\newblock High-performance line bundle cohomology computation based on
  \cite{Blumenhagen:2010pv}
  \href{http://wwwth.mppmu.mpg.de/members/blumenha/cohomcalg/}{http://wwwth.mppmu.mpg.de/members/blumenha/cohomcalg/}.

\end{thebibliography}
\end{document}